\documentclass[fleqn,usenatbib]{mnras}

\usepackage{newtxtext,newtxmath}

\usepackage[T1]{fontenc}

\DeclareRobustCommand{\VAN}[3]{#2}
\let\VANthebibliography\thebibliography
\def\thebibliography{\DeclareRobustCommand{\VAN}[3]{##3}\VANthebibliography}

\usepackage{graphicx}	
\usepackage{amsmath}	

\usepackage{epstopdf}
\title[Detailed analysis of the wide binary HD 202772A/B]{Detailed spectroscopic and photometric analysis of the remarkable planet-hosting wide binary system HD 202772A/B}

\author[E. Jofré et al.]{
Emiliano Jofré $^{1,2}$\thanks{E-mail: emiliano.jofre@unc.edu.ar},
Yilen Gómez Maqueo Chew $^{3}$, 
Romina Petrucci $^{1,2}$,         
Carlos Saffe $^{2,4,5}$,
\newauthor
Jhon Yana Galarza $^{6,15}$,
Paula Miquelarena $^{2,4,5}$,
María Paula Ronco $^{2,7,8}$,
Matías Cerioni $^{2,9}$,
\newauthor
Camila Zuloaga $^{1,2}$,
Eder Martioli $^{10}$,
Francesca Faedi $^{11}$,
Cintia Martínez $^{1,2}$,
Leila Saker $^{1}$,
\newauthor
Jorge Meléndez $^{12}$,
Matías Flores Trivigno $^{2,4,5}$,
Leslie Hebb $^{13}$,
Rodrigo Díaz $^{2,14}$,
and Luciano García $^{1}$
\\
$^{1}$ Universidad Nacional de C\'ordoba - Observatorio Astron\'{o}mico de C\'{o}rdoba, Laprida 854, X5000BGR, C\'ordoba, Argentina\\
$^{2}$ Consejo Nacional de Investigaciones Científicas y Técnicas (CONICET), Godoy Cruz 2290, CPC 1425FQB, CABA, Argentina\\
$^{3}$ Instituto de Astronom\'ia, Universidad Nacional Aut\'onoma de M\'exico, Ciudad Universitaria, 04510 Ciudad de M\'exico, M\'exico \\
$^{4}$  Instituto de Ciencias Astronómicas, de la Tierra y del Espacio (ICATE), España Sur 1512, CC 49, 5400 San Juan, Argentina\\
$^{5}$Facultad de Ciencias Exactas, Físicas y Naturales, Universidad Nacional de San Juan, San Juan, Argentina\\
$^{6}$The Observatories of the Carnegie Institution for Science, 813 Santa Barbara Street, Pasadena, CA 91101, USA\\
$^{7}$ Facultad de Ciencias Astronómicas y Geofísicas, Universidad Nacional de La Plata, Paseo del Bosque s/n, 1900 La Plata, Argentina\\
$^{8}$ Instituto de Astrofísica de La Plata (IALP), CCT La Plata-CONICET-UNLP, Paseo del Bosque s/n, La Plata, Argentina\\
$^{9}$ Instituto de Astronomía Teórica y Experimental (IATE), Laprida 854, (X5000BGR) Córdoba, Argentina\\
$^{10}$ Laboratório Nacional de Astrofísica, Rua Estados Unidos 154, Itajubá, MG, 37504-364, Brazil\\
$^{11}$ School of Physics and Astronomy, University of Leicester, University Road, Leicester LE1 7RH, UK  \\
$^{12}$ Departamento de Astronomia do IAG/USP, Universidade de São Paulo, Rua do Matão 1226, 05508-090 São Paulo, SP, Brazil\\
$^{13}$ Department of Physics, Hobart and William Smith Colleges, 300 Pulteney Street, Geneva, NY 14456, USA\\
$^{14}$ International Center for Advanced Studies, UNSAM--CONICET, Argentina\\
$^{15}$Departamento de Astronomía, Facultad de Ciencias Físicas y Matemáticas Universidad de Concepción, Av. Esteban Iturra, Casilla 160-C, Chile
}

\date{Accepted 2025 October 29. Received 2025 October 21; in original form August 2}

\pubyear{\the\year{}}

\begin{document}
\label{firstpage}
\pagerange{\pageref{firstpage}--\pageref{lastpage}}
\maketitle

\begin{abstract}
We conducted a detailed spectroscopic and photometric characterization of the planet-hosting wide binary HD 202772A/B. No planet has been detected around HD 202772B, whereas HD 202772A, more evolved than its companion and near the end of its main-sequence (MS) phase, hosts a transiting hot Jupiter. The system has one of the hottest components ($T_{\mathrm{eff}; A} \sim 6440$ K) and one of the largest surface gravity differences between components ($\Delta\log g_{A-B} \sim 0.4$ dex) among MS planet-hosting wide binaries. Using a global fit including our stellar parameters, radial velocities, and new TESS data, we derive refined properties of the planet orbiting HD 202772A, finding it to be the most irradiated hot Jupiter known in a wide binary. We also constrain the presence of additional transiting planets around HD 202772A and new transiting planets around HD 202772B using TESS photometry. We derive high-precision, strictly differential abundances for 27 species based on Gemini-GRACES spectra. HD 202772A shows lower lithium abundance (by 0.45 dex) relative to B, consistent with their stellar parameter differences. We also detect a small but significant enhancement in refractory elements in HD 202772A, particularly those with condensation temperatures above 1400 K (+0.018 $\pm$ 0.004 dex). We explored several scenarios to explain the observed chemical anomalies. Our analysis suggests that rocky planet engulfment, primordial inhomogeneities, and $\delta$ Scuti-related effects are unlikely to fully account for the chemical pattern. Instead, the differences observed in certain refractory elements might support atomic diffusion as the most plausible explanation.
\end{abstract}




\section{Introduction}

In addition to the well-established correlation between stellar metallicity  and the occurrence rate of short-period giant planets \citep[e.g.,][]{Gonzalez1997, Fischer2005, Ghezzi2010}, the analysis of detailed and high precision differential chemical abundances of several elements has enabled the discovery of other subtle anomalous chemical patterns that also might be connected to the planet formation processes and the subsequent planetary evolution. \citet[][hereafter M09]{Melendez2009} found that the Sun is depleted in refractory elements (T$_{c}$ $\gtrsim$ 1000 K) relative to volatiles (T$_{c}$ $<$ 1000 K) when compared to a sample of  11 nearby solar twins, and that the abundance differences strongly correlate with T$_{c}$ \citep[see also][]{Ramirez2009, Bedell2018, Nibauer2021}. M09 proposed that the missing refractories in the solar atmosphere could be locked in the rocky planets, asteroids, and rocky cores of giant planets in the solar system.  More recently, it has been suggested that T$_{c}$ trends can also be caused by the gap created by the formation of distant giant planets (i.e., Jupiter-analog) that prevents the accretion of refractories exterior to its orbit onto the star \citep{Booth2020, Huhn2023}. Alternatively, several stars show enhancement of refractory elements, relative to volatiles, when compared to field stars with similar atmospheric parameters \cite[e.g.,][]{Schuler2011, Melendez2017, Liu2020, Church2020}. It has been suggested that the dynamical evolution of planetary systems can lead to engulfment events \citep[e.g.,][]{Chatterjee2008, Carrera2019, Lau2025}. Additional support for planetary engulfments is also based on transient luminosity optical events, such as ZTF SLRN-2020, that could be powered by the accretion of a gas-giant planet \citep[e.g.,][]{Soker2023, De2023, Lau2025}. 

Alternatively, it has been debated that the observed T$_{c}$ correlations may arise from mechanisms unrelated to the presence of planets such as Galactic chemical evolution \citep[e.g.,][]{Adibekyan2014, Nissen2015} and/or radiative dust cleansing \citep{Onehag2011, Onehag2014, Gustafsson2018a, Gustafsson2018b}. One effective way to mitigate or eliminate these effects is by studying binary systems. Observations and numerical simulations suggest that the components of these systems form simultaneously from the same molecular cloud and share the same initial chemical composition \citep{Desidera2006, Kratter2011, King2012, Reipurth2012, Alves2019, Andrews2019, Hawkins2020, Nelson2021}. In particular, the analysis of planet-hosting binary systems is key to understanding how planet formation and evolution influence the stellar composition regardless of environmental or Galactic chemical evolution effects.

To date, the multi-element chemical pattern of $\sim$40 planet hosting binaries\footnote{Strictly speaking, the stellar pairs used in these binary studies are, by definition, comoving stars. Although the components in each pair exhibit consistent astrometry and nearly identical systemic radial velocities, this does not guarantee that they are gravitationally bound \citep[see e.g.,][]{Oh2017, Andrews2017, Ramirez2019}.} has been scrutinized \citep[e.g.,][]{Ramirez2011, Ramirez2015, Liu2014, Mack2014, Biazzo2015, Teske2016a, Teske2016b, Saffe2015, Saffe2017, Saffe2019, Tucci2019, Jofre2021, Spina2021, Behmard2023, Flores2024, YanaGalarza2024}. Due to the variety of results and interpretations from these studies, it is still inconclusive whether planet formation and/or evolution leave a detectable signature in the observed differential chemical pattern of wide binaries. Moreover, recent studies of binaries have raised the possibility that the origin of the chemical anomalies could be due to primordial inhomogeneities in the molecular clouds from which the stars in each binary formed \citep{Behmard2023, Saffe2024, Soliman2025}. Furthermore, the precise chemical analysis of binaries, especially those with large differences between the surface gravities of their components, can reveal the impact of atomic diffusion on their abundance pattern \citep{Liu2021}. Nevertheless, up to now, the peculiar abundance pattern of only 4 planet-hosting wide binary systems may be attributed to the effects of atomic diffusion \citep{Liu2021}.

Of the total sample of planet-hosting binaries with abundance analyses, only 12 of them correspond to main-sequence (MS) pairs with planets exclusively detected around one component. Moreover, just for this small sample, the precise abundances are based on the strictly line-by-line differential technique using high signal-to-noise ratio \citep[SNR $\gtrsim$ 200; e.g.,][]{Ramirez2011, Liu2014, Mack2014, Saffe2015, Saffe2017, Saffe2019, Tucci2019, Jofre2021, Flores2024, YanaGalarza2024}\footnote{The other binary systems host at least one planet around each component. \citep[e.g.,][]{Mack2014, Biazzo2015, Ramirez2015, Teske2016a, Teske2016b}.}. 
In this work we expand the small sample of planet-hosting binaries by investigating, for the first time, the chemical composition of the HD 202772A/B binary system via the strictly differential technique from high-quality spectra. A transiting hot Jupiter planet ($M_{\mathrm{p}}$ $\sim$1 $M_{\mathrm{J}}$) around HD 202772A was discovered by the TESS mission \citep[][hereafter W19]{Wang2019}. No planet has been reported yet around HD 202772B. Considering that the components are separated by $\sim$1.5'' (projected separation of only $\sim$200 au), and have consistent systemic RVs, proper motions, and parallaxes, W19 suggested that the two stars are likely gravitationally bound. Also, using the Besan\c{c}on Galactic Model \citep{Robin2003}, W19 estimated that the probability of lucky alignment of the two stars is negligible\footnote{Moreover, following \citet{Andrews2017} and \citet{Ramirez2019}, we examined the location of the HD 202772A/B pair in the plot of projected separation versus total velocity difference. With a projected separation of s = 200 au between the components ($\log s$ = 2.30 au) and a difference in total velocity of $\Delta V$ = 0.72 km $s^{-1}$ ($\log \Delta V \sim$ $-$0.14 km $s^{-1}$), HD202772 A/B fall safely within the limits where a system can be considered as physically bound and hence a true binary.}. One of the remarkable properties of this binary is that HD 202772A is considerably more evolved than B ($\Delta \log g \sim$ 0.4 dex,  which is one of the largest differences among the MS wide binaries analyzed so far) and near the end of its MS life (W19). Thus, this system is also an ideal laboratory for exploring the effects of atomic diffusion on stellar abundances and studying the radius inflation of the hot Jupiter HD 202772A b.

To our knowledge, the only previous study dealing with the detailed multi-element chemical pattern of this system was carried out by \citet[][hereafter BE23]{Behmard2023}. In that work, BE23 studied a sample of 36 planet-hosting binary systems to test the planet engulfment scenario. BE23 did not find signatures of planet engulfment in HD 202772A/B. However, in the present work we propose to use higher-quality spectra together with a fully differential determination of stellar parameters and abundances.


Using our derived high-precision stellar parameters, literature radial velocity data, and additional new TESS sectors photometry, we also computed refined physical and orbital properties of the hot Jupiter HD 202772A b. This information, along with the precise chemical stellar abundances presented here, will be crucial for analyzing and interpreting future measurements with JWST \citep[e.g.,][]{Adams2024}. 

In Section \ref{sec.observations}, we describe our spectroscopic and photometric observations. In Section \ref{sec.stellar}, we present the stellar characterization analysis that includes the computation of stellar parameters and high-precision elemental abundances for 27 species. In Section \ref{sec.differences.abundances}, we investigate the chemical differences between HD 202772A and HD 202772B, while in Section \ref{sec.planetary} we present the refined planetary parameters. In Section \ref{sec.discussion} we discuss some scenarios that could explain the observed chemical pattern and also the planetary parameters of HD 202772A b in the context of radius inflation. Finally, in Section \ref{sec.conclusions} we present our main conclusions.

\section{Observations and data reduction} \label{sec.observations}

\subsection{High-resolution Spectroscopy}
We observed HD 202772A and HD 202772B on July 2020 with the Gemini Remote Access to CFHT ESPaDOnS Spectrograph \citep[GRACES;][]{Chene2014} at the 8.1-m Gemini North telescope. Observations were carried out in  queue mode (GN-2020A-Q-126, PI: E. Jofr\'e) in the one-fiber mode (object only), which achieves a resolving power of R $\sim$ 67,500 between 400 and 1,050 nm. We obtained consecutive exposures of 4 $\times$ 190 s for the A component and 3 $\times$ 950 s for the B star. 

These observations, along with a series of calibrations (including 6 $\times$ ThAr arc lamp, 10 $\times$ bias, and 7 $\times$ flat-field exposures), were used as input in the OPERA (Open source Pipeline for ESPaDOnS Reduction and Analysis) code\footnote{OPERA is available at \url{http://wiki.lna.br/wiki/espectro}.} software \citep{Martioli2012} to obtain the reduced data. The details of the reduction can be found in \citet{Jofre2020}. Briefly, the reduction includes optimal extraction of the orders, wavelength calibration, and normalization of the spectra.  

Using the IRAF\footnote{IRAF is distributed by the National Optical Astronomy Observatories, which are operated by the Association of Universities for Research in Astronomy, Inc., under cooperative agreement with the National Science Foundation.} task \texttt{scombine}, the individual exposures of each target were co-added to obtain the final spectra which present a signal-to-noise ratio per resolution element of S/N $\sim$ 500 and S/N $\sim$ 450 around 6000 \AA, for the A and B stars, respectively. In addition, for the spectroscopic analysis described below, we used a solar spectrum (reflected sunlight from the Moon) obtained with the same GRACES setup (SNR $\sim$ 410 at 6000 \AA). 

\subsection{Photometric variability}\label{photvar}

We assessed the short- and long-term photometric variability of the system through the analysis of data obtained with different space- and ground-based facilities. 

\subsubsection{TESS}
HD 202772A/B was observed by the space mission TESS \citep{Ricker2015} at 2-min cadence in sector 1 (from 25 July to 22 August 2018) and in sectors 28 and 68 (between 31 July and 25 August 2020 and from 29 July to 25 August 2023) in 20-sec and 2-min cadences. Both stars fall within the same TESS pixel (Figure \ref{fig.TPF}), hence it is not possible to separate the light curves of each component. We downloaded the Presearch Data Conditioning Simple Aperture Photometry (PDCSAP), processed with the TESS SPOC pipeline \citep{jenkins2016}. Based on the ephemeris and the planetary parameters determined in Section \ref{exofastv2}, we eliminated all the in-transit data-points from the light curve. Also, we discarded points with evident systematics at the start, end, and between the gap of some sectors that were not entirely removed by the SPOC pipeline. The precision in the resulting TESS light curves is 1 mmag at the 20-sec cadence and 0.47 mmag at the 2-min cadence.

\subsubsection{LCOGT}
We observed HD 202772A/B in the i'-band on the night of July 6, 2023, using the ground-based LCOGT 1m telescope at CTIO (Chile) equipped with a SINISTRO camera (NSF2023A-007, PI: Leslie Hebb). The images were automatically processed with the LCO BANZAI pipeline \citep{xu2021} and the data were extracted using aperture photometry with the AstroImageJ (AIJ) software package \citep{collins2017}. Some of the data points showing anomalous flux values were eliminated. In total, we finally used 680 observations with 19.3 mmag precision in the i'-band. In this case, neither was possible to separate both components in the images. 

\subsubsection{ASAS}
We used V-band observations taken between 2001 and 2009 from the All Sky Automated Survey \citep[ASAS,][]{pojmanski2002}. We only employed the magnitudes flagged as ``A'' or ``B''\footnote{Each ASAS observation is classified according to its quality in four grades, being ``A'' the best data, ``B'' mean data, ``C'' data with no measured magnitude, and ``D'' the worst data.}. Available ASAS magnitudes are computed with five different apertures, where the smallest one has a size of 30''. Given that the stars in the system are separated by 1.3'', none of the apertures allow to resolve the components. Then, we decided to adopt the V-magnitudes calculated as the weighted mean of the magnitudes measured with each aperture. After removing some bad points, we kept a total of 377 observations with precision in V of 15 mmag. The planetary transit is not visible when phase-folding the data with the ephemeris computed in Section \ref{exofastv2}. This, along with the fact that the errors in magnitude are larger than the transit depth, made it unnecessary to eliminate the in-transit data-points.

\subsubsection{Tycho}
HD 202772A/B was observed by the ESA Hipparcos satellite from 1990 to 1992. We extracted the combined magnitude of both stars in the system from the Tycho-2 catalogue \citep{hog2000}, which is based on the $V_{T}$- and $B_{T}$-band observations collected by the star mapper of the mission. We excluded outliers and magnitudes without error measurements. For this reason, we only used 316 and 312 photometric data-points in the $V_{T}$ and $B_{T}$ filters, respectively. The light curves present a dispersion of $\sigma_{V}=$ 0.7 and $\sigma_{B}=$0.64 mag. As for the ASAS observations, the planetary transit was not detected in the phase-folded data, hence there was no need to remove in-transit data-points.\\

Time-series data were analyzed with the tools available at the LIGHTKURVE Python package \citep{Lightkurve}. We searched for periodic modulations in all the light curves following the criteria described in \citet{petrucci2024} and also sought for flares with the \texttt{altaipony} code \citep{davenport2016, ilin2021}. No evidence for a short rotation period, prolonged photometric activity cycle, or flaring events was found. Nonetheless, we note that all these light curves correspond to the combined flux of both components. Light curves of each individual star are needed to confirm the absence of photometric variability. 


\section{Analysis} \label{sec.stellar}
\subsection{Atmospheric stellar parameters} \label{sec.atmospheric.parameters}
We computed the atmospheric parameters T$_{\mathrm{eff}}$ (effective temperature), $\log g$ (surface gravity), [Fe/H] (abundance of iron), and $v_{micro}$ (microturbulent velocity) from the equivalent widths (EWs) of iron lines by imposing the differential spectroscopic equilibrium technique \citep[see details in e.g.,][]{Saffe2017, Ramirez2019, Jofre2021}. Briefly, the T$_{\mathrm{eff}}$ and $v_{micro}$ are iteratively modified until the correlation of differential iron abundance (Fe \textsc{i} lines only) with excitation potential (EP = $\chi$) and reduced equivalent width (REW = $\log \mathrm{EW}/\lambda$), respectively, are minimized (Figure \ref{fig.equilibrium}). Simultaneously, $\log g$ is modified until the differential abundance derived from Fe \textsc{i} lines approaches the one computed from Fe \textsc{ii} lines. To carry out this process automatically, as in our previous works, we used the \texttt{Qoyllur-Quipu} (or \texttt{q$^{2}$}) Python package\footnote{The code is available at \url{https://github.com/astroChasqui/q2}} \citep{Ramirez2014}, which uses the \texttt{MOOG} software \citep{Sneden1973}. Also, we configured \texttt{q$^{2}$} to employ linearly interpolated 1D local thermodynamic equilibrium (LTE) Kurucz ODFNEW model atmospheres \citep{Castelli2003}. The iron line list and the atomic parameters (EP and oscillator strengths, $\log g_f$) were taken from \citet{Ramirez2015} and the EWs were manually measured using the \texttt{splot} task in IRAF. 

%
%
%

\begin{table}
\caption{Atmospheric parameters of HD 202772A and HD 202772B}
\label{table.spectroscopic.parameters}
\centering
\begin{tabular}{l c c c}
\hline\hline
 &	HD 202772A 	&	HD 202772B 	&	$\Delta$(A$-$B) \\ 	
\hline
\hline
\multicolumn{4}{c}{\citet{Wang2019}} \\	
\hline	
$T_{\mathrm{eff}}$ 	&	6330 $\pm$ 100	&	6156 $\pm$ 100	&	174 $\pm$ 141		\\
$\log g$ 	&	4.03 $\pm$ 0.10	&	4.24 $\pm$ 0.10	&	$-$0.21 $\pm$ 0.14 	\\
$\mathrm{[Fe/H]}$ 	&	0.29 $\pm$ 0.06	&	0.25 $\pm$ 0.06	&	0.04 $\pm$ 0.08		\\
$v_{\mathrm{micro}}$ 	&	n/a			&	n/a			&	n/a			\\
\hline
\multicolumn{4}{c}{\citet{Flores2022}} \\	
\hline	
$T_{\mathrm{eff}}$	&	6484 $\pm$ 24	&	6390 $\pm$ 20	&	94 $\pm$ 31		\\
$\log g$ 	&	4.08 $\pm$ 0.04 &	4.36 $\pm$ 0.04 &	$-$0.28 $\pm$ 0.06 	\\
$\mathrm{[Fe/H]}$ 	&	0.203 $\pm$ 0.015	&	0.263 $\pm$ 0.014	&	$-$0.06 $\pm$ 0.02		\\
$v_{\mathrm{micro}}$ 	&	2.12 $\pm$ 0.04	&	1.83 $\pm$ 0.03			&	0.29 $\pm$ 0.05			\\
		
\hline
\multicolumn{4}{c}{\citet{Behmard2023}} \\	
\hline	
$T_{\mathrm{eff}}$ 	&	6255 $\pm$ n/a	& 6103	 $\pm$ n/a	&	152 $\pm$ n/a		\\
$\log g$ 	&	3.91 $\pm$ n/a	&	4.14 $\pm$ n/a	&	$-$0.23 $\pm$ n/a 	\\
$\mathrm{[Fe/H]}$ 	&	0.27 $\pm$ n/a &	0.25 $\pm$ n/a	&	0.02 $\pm$ n/a		\\
$v_{\mathrm{micro}}$ 	&	n/a			&	n/a			&	n/a			\\
\hline
\multicolumn{4}{c}{This work, solar reference (Section \ref{sec.atmospheric.parameters})}\\							
\hline
$T_{\mathrm{eff}}$ 	&	6443 $\pm$ 26		&	6245 $\pm$ 22		&	198 $\pm$ 34		\\
$\log g$ 		&	3.94 $\pm$ 0.08		&	4.33 $\pm$ 0.06		&	$-$0.39 $\pm$ 0.10	\\
$\mathrm{[Fe/H]}$ 	&	0.229 $\pm$ 0.02	&	0.229 $\pm$ 0.02	&	0.00 $\pm$ 0.06		\\
$v_{\mathrm{micro}}$ 	&	1.83 $\pm$ 0.04		&	1.39 $\pm$ 0.04		&	0.44 $\pm$ 0.06		\\
\hline
\multicolumn{4}{c}{This work, HD 202772B reference (Section \ref{sec.atmospheric.parameters})}\\
\hline
$T_{\mathrm{eff}}$ &	6442 $\pm$ 10		&	6245 $\pm$ 22		&	197 $\pm$ 24		\\
$\log g$ 		&	3.93 $\pm$ 0.03		&	4.33 $\pm$ 0.06		&	$-$0.40 $\pm$ 0.07	\\
$\mathrm{[Fe/H]}$ 	&	0.232 $\pm$ 0.007	&	0.229 $\pm$ 0.02	&	0.003 $\pm$ 0.021	\\
$v_{\mathrm{micro}}$ 	&	1.83 $\pm$ 0.02		&	1.39 $\pm$ 0.04		&	0.44 $\pm$ 0.04		\\
\hline
\end{tabular}
\vskip 0.05 cm
\noindent {\footnotesize{Note: The units for the fundamental parameters are: $T_{\mathrm{eff}}$ (K), $\log g$ (dex). $\mathrm{[Fe/H]}$  (dex), and $v_{\mathrm{micro}}$ (km $s^{-1}$). }\\
}
\end{table}

The results are summarized in Table \ref{table.spectroscopic.parameters}, where the first three blocks show the literature values. In particular, the ones obtained by \citet{Wang2019} were  employed  in our analysis as initial values. The fourth block includes the parameters resulting from the line-by-line differential analysis using our solar spectrum as reference, adopting as solar parameters T$_{\mathrm{eff}}$ = 5777 K, $\log g$ = 4.44 dex, [Fe/H] = 0 dex, and $v_{micro}$ = 1.0 km s$^{\mathrm{-1}}$. Finally, the last block lists the stellar parameters of HD 202772A but considering the HD 202772B as reference by adopting its solar-reference parameters derived above as fixed. Figure \ref{fig.equilibrium} shows the spectroscopic equilibrium for the (A $-$ B) final result. 

As can be seen in Table \ref{table.spectroscopic.parameters}, HD 202772A is  $\sim$ 200 K hotter than HD 202772B, and both components show a negligible difference in [Fe/H] (i.e., $\Delta$[Fe/H] = 0.003 dex). However, the stars exhibit one of the largest differences in the surface gravities ($\Delta$$\log g$ $\sim$ 0.4 dex) among MS planet-hosting wide binaries which, as it is discussed in Sec. \ref{sec.atomic.diffusion}, allows us to examine the effects of atomic diffusion. As can be noticed in Figure \ref{fig.isochrones}, the planet host star HD 202772A is more evolved than HD 202772B but remains within the MS limits according to physically motivated boundaries from solar metallicity interior models \citep{Huber2017, Berger2018} using the \texttt{evolstate} code\footnote{See \url{https://github.com/danxhuber/evolstate}}. Using MIST models \citep{Choi2016}, we also estimated that HD 202772A will exhaust its hydrogen fuel, and leave its MS life, in $\sim$0.4 Gyr which agrees with the results from \citet{Wang2019}.



 \begin{table*}
  \small
      \caption{Stellar parameters of HD 202772A and HD 202772B}
         \label{table.all.stellar.parameters}
     \centering
         \begin{tabular}{l c c c}
            \hline
\hline            
Parameter	&	HD 202772A			&	HD 202772B	& Source	\\
										
\hline	
\multicolumn{4}{c}{Adopted atmospheric and physical  parameters} \\											
\hline											
Effective temperature, $T_{\mathrm{eff}}$ (K)	&	6442	$\pm$	10	&	6245	$\pm$	22	&	This work (Sec. \ref{sec.atmospheric.parameters})	\\
Surface gravity, $\log g$ (cgs) 	&	3.93	$\pm$	0.03	&	4.33	$\pm$	0.06	&	This work (Sec. \ref{sec.atmospheric.parameters})	\\
Metallicity, $\mathrm{[Fe/H]}$ (dex) 	&	0.232	$\pm$	0.01	&	0.229	$\pm$	0.02	&	This work (Sec. \ref{sec.atmospheric.parameters})	\\
Microturbulent velocity, $v_{\mathrm{micro}}$ (km $s^{-1}$)	&	1.83	$\pm$	0.02	&	1.39	$\pm$	0.04	&	This work  (Sec. \ref{sec.atmospheric.parameters})	\\
Macroturbulent velocity, $v_{\mathrm{macro}}$ (km $s^{-1}$)	&	7.27	$\pm$	0.23	&	5.49	$\pm$	0.23	&	This work (Sec. \ref{sec.abundances})	\\
Rotational velocity, $v\sin{i}$ (km $s^{-1}$)	&	4.75	$\pm$	0.25	&	3.10	$\pm$	0.52	&	This work (Sec. \ref{sec.abundances})	\\
Mass, $M_{\mathrm{\star}}$ ($M_{\mathrm{\odot}}$)	&	$1.721^{+0.054}_{-0.058}$	&	$1.197^{+0.044}_{-0.054}$	&	This work (Sec. \ref{sec.other.parameters} \& \ref{exofastv2})	\\
Radius, $R_{\mathrm{\star}}$ ($R_{\mathrm{\odot}}$)	&	$2.554^{+0.056}_{-0.052}$	&	$1.175\pm 0.028$ &	This work (Sec. \ref{sec.other.parameters} \& \ref{exofastv2})\\
Age, $\tau_{\mathrm{\star}}$ (Gyr)	&	$1.63^{+0.27}_{-0.21}$	&	$1.32^{+1.6}_{-0.93}$ &	This work (Sec. \ref{sec.other.parameters} \& \ref{exofastv2})\\
Luminosity, $L_{\mathrm{\star}}$ ($L_{\mathrm{\odot}}$)	&	$10.22^{+0.47}_{-0.41}$&  1.881	$\pm$ 0.080 &	This work (Sec. \ref{sec.other.parameters} \& \ref{exofastv2}) \\
\hline											
\multicolumn{4}{c}{Astrometry and kinematics} \\											
\hline											
Proper motion in RA, $\mu_{\alpha}$ (mas yr$^{-1}$)	& 23.25	$\pm$ 0.04		&	28.91	$\pm$ 0.07		&	\textit{Gaia} DR3	\\
Proper motion in DEC, $\mu_{\delta}$ (mas yr$^{-1}$)	 &$-$57.67 $\pm$ 0.03	&	-56.51	$\pm$ 0.08		&	\textit{Gaia} DR3	\\
Parallax, $\pi$ (mas)	&	6.14	$\pm$ 0.04		&	6.33 $\pm$ 0.06	&	\textit{Gaia} DR3	\\
Barycentric radial velocity, RV (km $s^{-1}$)	&	$-$17.97	$\pm$	0.05	&	$-$17.25	$\pm$	0.07	&	This work (Sec. \ref{sec.observations})	\\
Space velocity component, U (km $s^{-1}$)	&	5.03	$\pm$	0.19	&	8.10	$\pm$	0.20	&	This work (Sec. \ref{sec.abundances})	\\
Space velocity component, V (km $s^{-1}$)	&	$-$42.77	$\pm$	0.32	&	$-$40.72	$\pm$	0.20	&	This work (Sec. \ref{sec.abundances})	\\
Space velocity component, W (km $s^{-1}$)	&	$-$2.12	$\pm$	0.25	&	$-$4.33	$\pm$	0.29	&	This work (Sec. \ref{sec.abundances})	\\
Galaxy population membership	&	Thin  disk			&	Thin  disk			&	This work (Sec. \ref{sec.abundances})	\\
\hline											
  \end{tabular} 
     
        \end{table*}

\subsection{Stellar masses, radii, and ages} \label{sec.other.parameters}


\begin{figure}
\centering
\includegraphics[width=.5\textwidth]{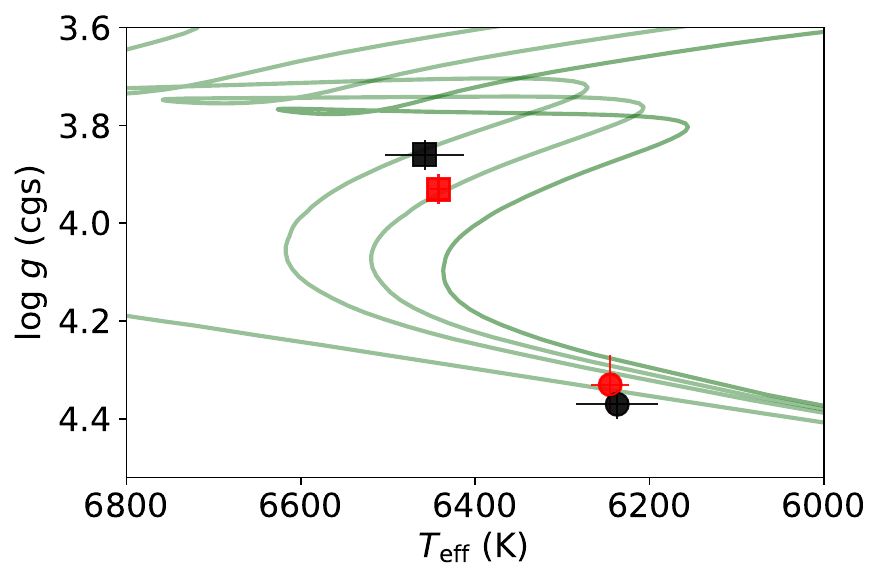 }
\caption{Location of HD 202772A (squares) and HD 202772B (circles) in the T$_{\mathrm{eff}}$--$\log g$ diagram. The red and black symbols indicates the parameters derived in this work. In particular, the black ones represent the T$_{\mathrm{eff}}$ and $\log g$ values derived from the global analysis performed with EXOFASTv2 whilst those the in red indicate the adopted differential spectroscopic parameters. The solid lines mark MIST isochrones corresponding to ages of 1.0, 1.6, 1.8, and 2.0 Gyr (left to right) for [Fe/H] = 0.23 dex.
\label{fig.isochrones}}
\end{figure}

We derived the stellar mass M$_{\mathrm{\star}}$, radius R$_{\mathrm{\star}}$, and ages $\tau_{\mathrm{\star}}$ of both stars in the wide binary using  \texttt{EXOFASTv2}\footnote{\url{https://github.com/jdeast/EXOFASTv2}} \citep{Eastman2013, Eastman2019}. 
We used our newly derived effective temperature and metallicity from Table~\ref{table.spectroscopic.parameters}, the Spectral Energy Distribution (SED) constructed from the catalog broadband photometry and W19 with measurements in which both stars are resolved, the \textit{Gaia} DR3 parallax,  
and extinction in the direction of HD 202772. 
For both stars, we set Gaussian priors for the parameters for which we have independent constraints, namely the T$_{\mathrm{eff}}$, [Fe/H], and parallax, with widths corresponding to their 1$\sigma$ uncertainties. 
In the case of the planet host HD~202772A, we set an upper limit to the extinction of A$_V$ = 0.17236 \citep[W19,][]{Schlafly2011} to avoid unrealistic solutions and include all the available datasets, to incorporate the additional constraint on the stellar properties via the stellar density from the transit (see more details in Section~\ref{exofastv2}).
In the case of HD~202772B, we adopted the extinction value from the planet host solution A$_V$ = 0.062$^{+0.056}_{-0.043}$ with a Gaussian prior and then simultaneously fit a single-star SED and the MIST stellar models. 
The derived properties are reported in Table~\ref{table.all.stellar.parameters} for both stars and correspond to the median in the MCMC final distributions and the 1-$\sigma$ uncertainties to the 68\% confidence intervals. 
We note that the ages derived for both stars agree with being coeval within their uncertainties. 

Finally, \texttt{EXOFASTv2} also computes the effective temperatures from the analysis of the SED and trigonometric surface gravities. This allowed us to make a consistency check on the spectroscopic values. We obtained T$_{\mathrm{eff (SED)}}$= 6484 $\pm$ 87 K and  $\log g$$_{\mathrm{(trigonometric)}}$ = 3.86 $\pm$ 0.02 dex for component A, whilst for component B \texttt{EXOFASTv2} returned T$_{\mathrm{eff (SED)}}$= 6331 $\pm$ 91 K and  $\log g$$_{\mathrm{(trigonometric)}}$ = 4.37 $\pm$ 0.03 dex. In Figure \ref{fig.isochrones} we show these values in comparison with those derived from spectroscopy. As can be noticed, there is a relatively good agreement between both sets of parameters for both stars within their uncertainties.

\begin{figure}
\centering
\includegraphics[width=.45\textwidth]{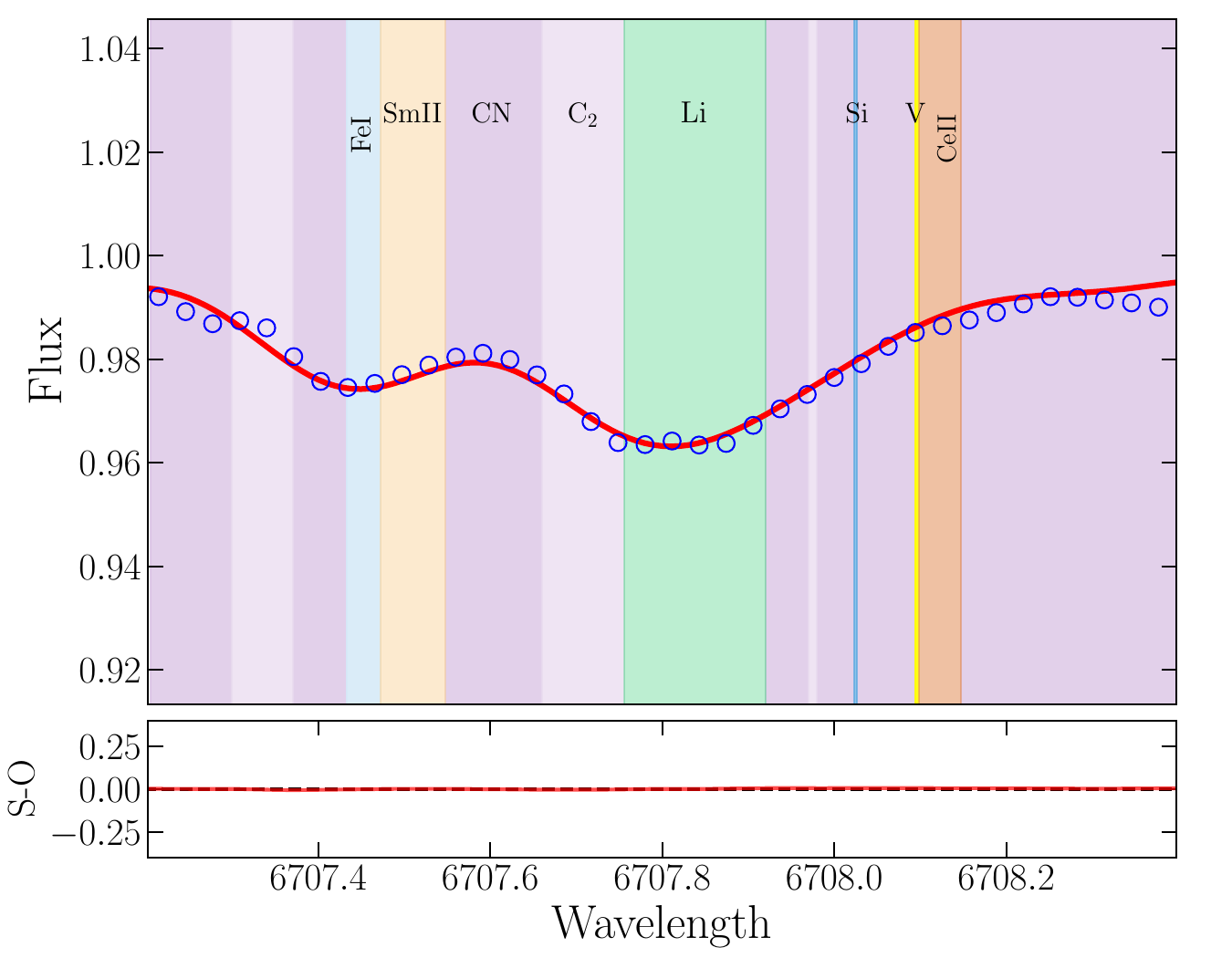}
\includegraphics[width=.45\textwidth]{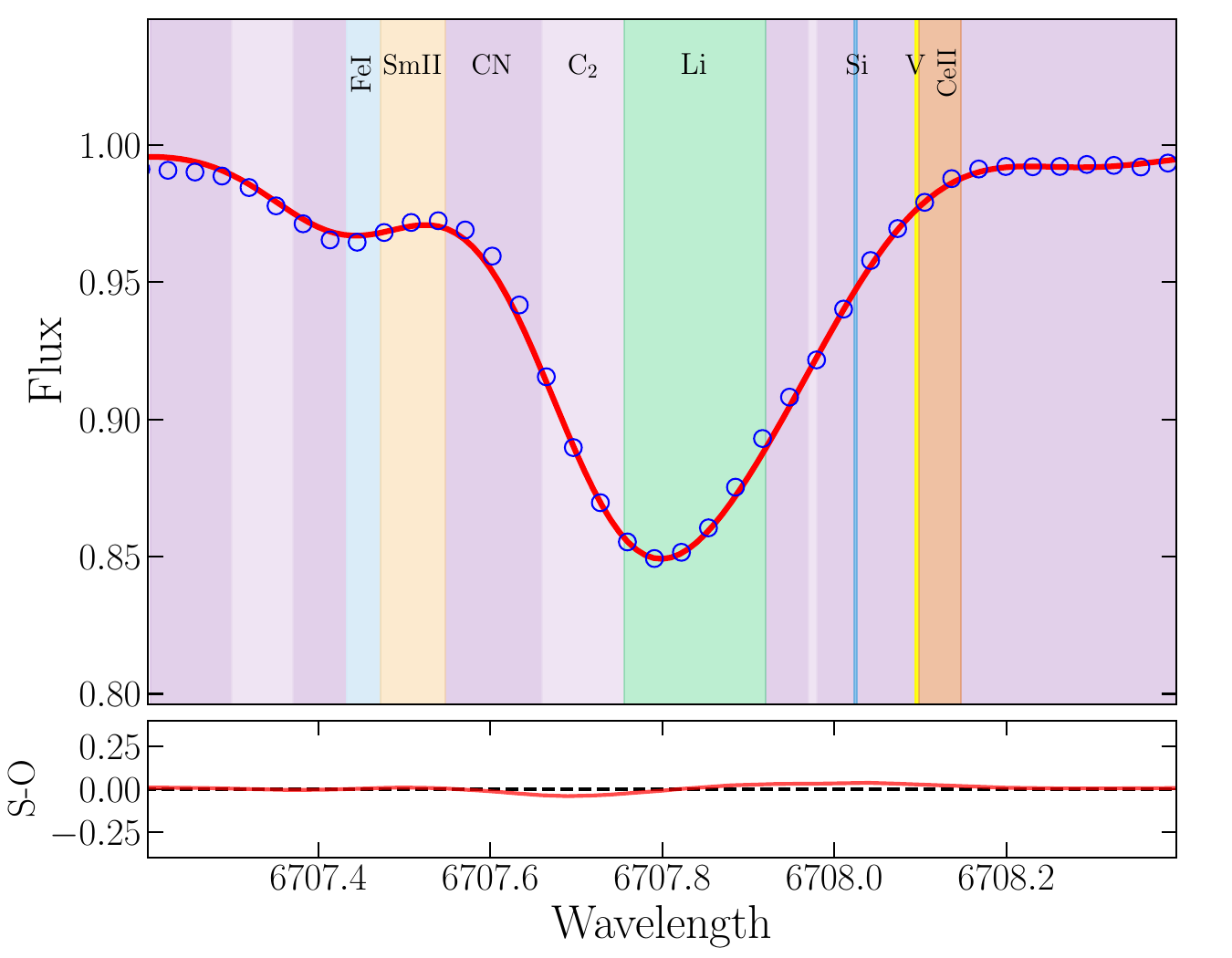}
\caption{Best fit obtained between the synthetic (red line) and the observed (blue circles) GRACES spectra of HD 202772A (upper panel) and HD 202772B (bottom panel) around the 6707.8 {\AA} lithium line. The colored vertical bands indicate the positions and contribution of different species in the region of the 6707.8 {\AA}  Li I line.
\label{fig.lithium.synth}}
\end{figure}

\subsection{Elemental abundances and galactic population} \label{sec.abundances}
We measured the abundances of 22 elements other than iron for the HD 202772 stars via equivalent widths and spectral synthesis, by adopting the spectroscopic atmospheric parameters presented in the last block of Table \ref{table.spectroscopic.parameters} as the first choice. The elemental abundances of C, O, Na, Mg, Al, Si, S, Ca, Sc, Ti, V, Cr, Mn, Co, Ni, Zn, Sr, Y, Zr, and Ba were derived from EWs using the curve-of-growth approach with the MOOG code (abfind driver) using the \texttt{q$^{2}$} code. The EWs were manually measured using the \texttt{splot} task in IRAF with the line list and atomic parameters adopted were taken from \citet{Jofre2021}. The abundances of Sc, Ti, and Cr (in addition to Fe) were obtained from both neutral and singly-ionized species, whilst for Y, Zr, Ba, and Ce only singly-ionized species are available. Hyperfine splitting was taken into account for V, Mn, Co, Y, and Ba from \citet[][]{Ramirez2014}. The O abundance was computed from the 7771-5 {\AA} IR triplet, adopting the non-LTE corrections by \citet{Ramirez2007}: $-$0.361 dex, $-$0.236 dex, and $-$0.128 dex for HD 202772A, HD 202772B, and the Sun, respectively. The oxygen abundances relative to the Sun and those differential are listed in Table \ref{table.abundances}.

\begin{table*}
 \small
\caption{Condensation temperatures, number of measured lines, and elemental abundances relative to the Sun ([X/H]) and differential between HD 202772A and HD 202772B ($\Delta$[X/H]$_{A-B}$).}
         \label{table.abundances}
     \centering
         \begin{tabular}{l c c c c c c c c}
            \hline\hline

Species	&	T$_{c}$ (K)	&	N$_{lines}$	&	[X/H] (dex)	&	Error (dex)	&	[X/H] (dex)	&	Error  (dex)	&	$\Delta$[X/H]$_{A-B}$ (dex)	&	Error (dex)	\\
	&		&		&	\multicolumn{2}{c}{(HD 202772A)}			&	\multicolumn{2}{c}{(HD 202772B)}			&		&		\\
\hline													

C	\textsc{i}	&	40	&	3	&	0.167	&	0.038	&	0.170	&	0.047	&	$-$0.003	&	0.010	\\
O	\textsc{i} $_{NLTE}$	&	180	&	3	&	0.073	&	0.016	&	0.071	&	0.021	&	0.009	&	0.012	\\
Na	\textsc{i}	&	958	&	5	&	0.328	&	0.031	&	0.335	&	0.044	&	0.012	&	0.022	\\
Mg	\textsc{i}	&	1336	&	4	&	0.153	&	0.088	&	0.148	&	0.054	&	0.010	&	0.009	\\
Al	\textsc{i}	&	1653	&	2	&	0.267	&	0.020	&	0.237	&	0.02	&	0.030	&	0.020	\\
Si	\textsc{i}	&	1310	&	23	&	0.264	&	0.040	&	0.229	&	0.038	&	0.035	&	0.005	\\
S	\textsc{i}	&	664	&	6	&	0.207	&	0.080	&	0.207	&	0.078	&	0.024	&	0.019	\\
Ca	\textsc{i}	&	1517	&	10	&	0.290	&	0.059	&	0.257	&	0.043	&	0.033	&	0.012	\\
Sc	\textsc{i}	&	1659	&	2	&	0.377	&	0.011	&	0.342	&	0.034	&	0.035	&	0.024	\\
Sc	\textsc{ii}	&	1659	&	8	&	0.265	&	0.062	&	0.258	&	0.086	&	0.036	&	0.023	\\
Ti	\textsc{i}	&	1582	&	34	&	0.275	&	0.067	&	0.235	&	0.063	&	0.038	&	0.010	\\
Ti	\textsc{ii}	&	1582	&	12	&	0.238	&	0.082	&	0.194	&	0.096	&	0.026	&	0.015	\\
V	\textsc{i}	&	1429	&	16	&	0.306	&	0.100	&	0.268	&	0.067	&	0.036	&	0.016	\\
Cr	\textsc{i}	&	1296	&	31	&	0.257	&	0.070	&	0.248	&	0.054	&	0.010	&	0.011	\\
Cr	\textsc{ii}	&	1296	&	4	&	0.200	&	0.084	&	0.151	&	0.086	&	$-$0.016	&	0.016	\\
Mn	\textsc{i}	&	1158	&	5	&	0.171	&	0.066	&	0.191	&	0.07	&	$-$0.020	&	0.031	\\
Fe	\textsc{i}	&	1334	&	144	&	0.230	&	0.008	&	0.227	&	0.005	&	0.003	&	0.007	\\
Fe	\textsc{ii}	&	1334	&	14	&	0.235	&	0.022	&	0.232	&	0.016	&	0.002	&	0.014	\\
Co	\textsc{i}	&	1352	&	6	&	0.245	&	0.053	&	0.248	&	0.06	&	$-$0.003	&	0.009	\\
Ni	\textsc{i}	&	1353	&	52	&	0.248	&	0.056	&	0.243	&	0.053	&	0.003	&	0.008	\\
Zn	\textsc{i}	&	726	&	3	&	0.116	&	0.060	&	0.129	&	0.034	&	$-$0.013	&	0.023	\\
Sr	\textsc{i}	&	1464	&	1	&	0.195	&	0.061	&	0.150	&	0.061	&	0.045	&	0.024	\\
Y	\textsc{ii}	&	1659	&	3	&	0.249	&	0.045	&	0.203	&	0.058	&	0.046	&	0.016	\\
Zr	\textsc{ii}	&	1741	&	1	&	0.284	&	0.061	&	0.249	&	0.061	&	0.035	&	0.028	\\
Ba	\textsc{ii}	&	1455	&	2	&	0.203	&	0.019	&	0.193	&	0.027	&	0.010	&	0.023	\\
Ce	\textsc{ii}	&	1478	&	1	&	0.063	&	0.037	&	0.033	&	0.040	&	0.030	&	0.054	\\
 Li	\textsc{i} $_{NLTE}$ $^{\textit{a}}$	&	1142	&	1	&	2.21	&	0.05	&	2.66	&	0.046	& $-$0.45		&	0.068	\\
\hline
\noalign{\smallskip}
\end{tabular} 
\vskip 0.05 cm
\noindent {\footnotesize{
$^{\textit{a}}$ Absolute abundances of lithium, A(Li).\\
}}

        \end{table*}


On the other hand, the abundance of Ce was obtained by performing a synthesis analysis using the synth driver of the MOOG code as detailed in \citet{Saffe2016}. 


We determined lithium abundance through spectral synthesis following the procedure given in \citet{Yana:2024arXiv241017590Y}. The method consists of first determining the solar macroturbulent velocity ($v_{\rm{macro}, \odot}$) using the solar spectra, a critical step for estimating the macroturbulent velocity ($v_{\rm{macro}}$) of the stellar pair. The $v_{\rm{macro}, \odot}$ was estimated via the spectral synthesis of five iron lines (6027.050, 6093.644, 6151.618, 6165.360, 6705.102 \AA) and one of nickel (6767.772 \AA) and fixing the solar projected rotational velocity ($v \sin i$) at 1.9 km s$^{-1}$ \citep{Saar:1997MNRAS.284..803S}. We obtained $v_{\rm{macro}, \odot} = 3.6 \pm 0.5$ km s$^{-1}$, consistent within the uncertainties with the value  of $3.2 \pm 0.5$ km s$^{-1}$ reported by \citet{Doyle:2014MNRAS.444.3592D} using the same instrument. Then, we computed the $v_{\rm{macro}}$ of the pair using the Eq. (1) of  \cite{Leo:2016A&A...592A.156D}. The $v \sin i$ of the system was calculated similarly to the solar $v_{\rm{macro}, \odot}$, but now fixing the estimated $v_{\rm{macro}}$. The estimated values of $v_{\rm{macro}}$ and $v \sin i$ for both stars are listed in Table \ref{table.all.stellar.parameters}.
Once both broadening velocities are obtained along with the instrumental broadening ($\lambda/R$) and the limb darkening coefficient (0.6), we synthesized the lithium line using the line list from \citet{Melendez:2012A&A...543A..29M} and the MOOG synth driver.

Figure \ref{fig.lithium.synth} shows the best fits to the Li features for both stars. We adopted the NLTE corrections of \citet{Lind2009}, which are 0.040 dex for star A and $-$0.004 dex for star B. Thus, the resulting abundances are A(Li)$_{NLTE}$ = 2.210 $\pm$ 0.05 dex and A(Li)$_{NLTE}$ = 2.660$^{+0.037}_{-0.046}$ dex for HD 202772A and HD 202772B, respectively.


The line-by-line differential abundances relative to the Sun ([X/H]) and those using HD 202772B as the reference star ($\Delta$[X/H]$_{A-B}$), along with the errors, are listed in Table \ref{table.abundances}. These errors are computed by adding in quadrature both the line-to-line scatter\footnote{For species with only one line (e.g., Zr \textsc{ii}, Sr \textsc{i}, and Ce \textsc{ii}), we conservatively adopted the largest line-to-line scatter obtained for species with more than three lines available.} and the errors propagated from each parameter uncertainty as in \citet{Ramirez2015}. Our errors in relative abundances, $\Delta$[X/H]$_{A-B}$, vary from one specie to another in the 0.005--0.031 dex range. In particular, the median error is 0.015 dex for volatiles and 0.014 for refractories. 


The $\alpha$-elemental abundances of HD 202772A ([$\alpha$/Fe]= 0.02 dex) and HD 202772B ([$\alpha$/Fe]= 0.0 dex) appear to be consistent with the  thin-disk kinematic membership of the system. The latter was computed following the procedure detailed in \citet{Jofre2015}. Briefly, using the radial velocities measured from the GRACES spectra and the proper motions and parallaxes from \textit{Gaia} DR3, we first derived the Galactic space-velocity components of each star (see second block of Table \ref{table.all.stellar.parameters}). Later, using these space-velocity values and the membership formulation by \citet{Reddy2006}, we found that the probability of belonging to the thin disk population is $\sim$98\% for both stars. 



\subsection{Comparison with previous studies}

The fundamental parameters of  HD 202772A and B were previously reported by W19, \citet[][F22 hereafter]{Flores2022}, and BE23\footnote{No errors for $\log g$ and $T_{\mathrm{eff}}$ are provided in BE23.}. All these previous values are summarized in Table \ref{table.spectroscopic.parameters}. Our estimations for both stars agree within the errors with those reported in the discovery paper by W19. However, with the exception of $\log g$ for the component A obtained by BE23 and F22 for the B star, we observe discrepancies above 1$\sigma$ for the other measurements. As noted before \citep{Hinkel2016, Jofre2017, Jofre2020}, the most probable sources for these discrepancies are the quality of the spectra (SNR and resolution), the use of different techniques employed (EWs vs. spectral synthesis), different line lists and model atmosphere. For example, BE23 determined the parameters from spectral synthesis using not so high SNR HIRES spectra (SNR = 141)\footnote{For reference, the spectral resolution of our Gemini-GRACES data is $\sim$25\% larger and, more importantly, the SNR of our spectra are $\sim$ 150\% and $\sim$ 250\% larger for HD 202772A and HD 202772B, respectively.}. In the case of the parameters derived by F22, although they also employed a differential analysis, the discrepancies might result from the use of lower-resolution data. In particular, their analysis was based on MIKE low-resolution spectra (R $\lesssim$ 35000; M. Flores, private communication).   

\begin{figure}
\centering
\includegraphics[width=.49\textwidth]{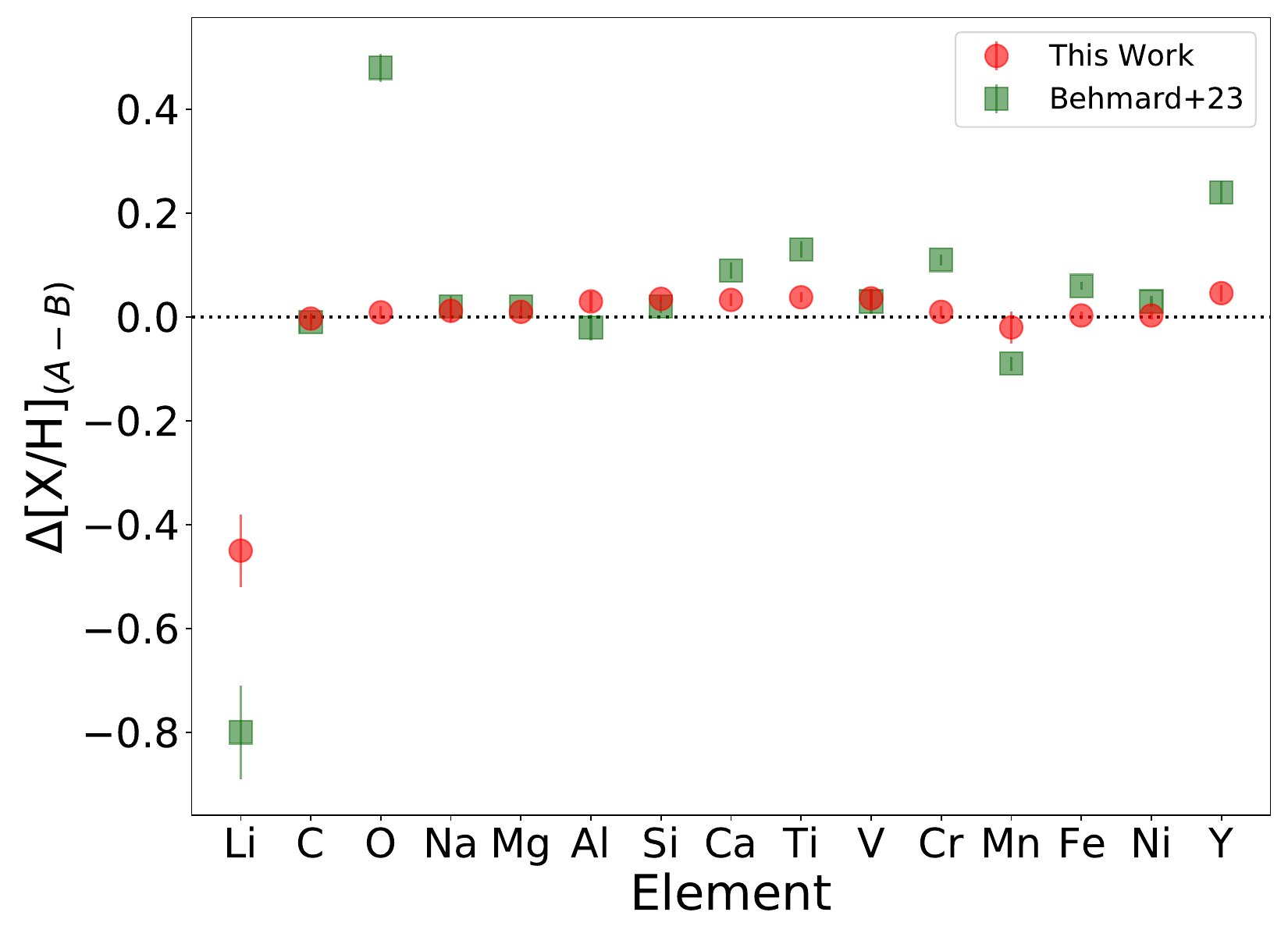}
\caption{Differential abundances (A$-$B) derived in this work (red circles) in comparison to those reported by BE23 (green squares) for common elements. The dotted line mark identical composition.\label{fig.comp.literature}}
\end{figure}

BE23 also derived multi-element chemical abundances of both stars. In Figure \ref{fig.comp.literature}
we show our abundances in comparison with those of BE23 for the elements in common. Although there is a relatively good agreement for some elements (C, Na, Mg, Si, V, Ni) the remaining elements show significant discrepancies (Li, O, Al, Ca, Ti, Cr, Mn, Fe, Y) with our results.
The causes of the disagreements are likely the same as previously mentioned for the differences with the atmospheric parameters from BE23. 

However, the significant discrepancy in oxygen ($\sim$0.5 dex) may also involve additional factors, such as a problem in the data reduction. In particular, since BE23 reported an abundance of [O/H] = 0.47 $\pm$ 0.02 dex and  [O/H] = $-$0.01 $\pm$ 0.02 dex, for HD 202772A and B, respectively, the discrepancy perhaps originated in the measurement obtained from the HIRES spectra of HD 202772A. 

Although BE23 do not explicitly mention whether they applied NLTE corrections to their oxygen abundances, we performed an additional check to explore whether the observed difference could be due to different non-LTE grids. To this end, we applied the 3D non-LTE correction grid from \citet{Amarsi2015} and obtained corrections of $-$0.168 dex, $-$0.142 dex, and $-$0.117 dex for components A, B, and the Sun, respectively. Taking these corrections into account, the resulting 3D NLTE oxygen abundances are: A(O)$_{NLTE; A}$ = 8.93 dex, A(O)$_{NLTE; B}$ = 8.832 dex, and A(O)$_{NLTE; Sun}$ = 8.68 dex. From these results, we derive a differential abundance of $\Delta$[O/H]$_{A-B}$ = 0.10 $\pm$ 0.012 dex, which is an order of magnitude larger than the one based on the NLTE corrections from \citet{Ramirez2007}. 
However, both estimates remain significantly lower than the $\sim$0.5 dex abundance difference reported in BE23. Therefore, the use of different NLTE correction grids alone does not account for the observed discrepancy.  

We also investigated whether the discrepancy in lithium abundances ($\sim$ 0.2 dex) could be attributed to the use of different NLTE grids. To this end, we employed the 3D NLTE correction grid from \citet{Wang2024}, via the \texttt{Breidablik} code\footnote{\url{https://github.com/ellawang44/Breidablik}}. Using this approach, we obtained A(Li)$_{3D-NLTE}$ = 2.17 dex for A and A(Li)$_{3D-NLTE}$ = 2.62 dex for B, which are nearly identical to those derived using the NLTE corrections from \citet{Lind2009}. Therefore, the discrepancies may arise from differences in methodology: while BE23 derived abundances from equivalent widths (EWs), we employed spectral synthesis.


Finally, the derived stellar radii and masses in this work are in good agreement (within 1$\sigma$) with literature values. The only exception corresponds to the mass of HD 202772A reported by B23 which is $\sim$17\% smaller than ours. The cause of this discrepancy is likely related to differences between our atmospheric parameters, especially for the $T_{\mathrm{eff}}$. 


\section{Chemical differences}\label{sec.differences.abundances}
\begin{figure*}
\centering
\includegraphics[width=.7\textwidth]{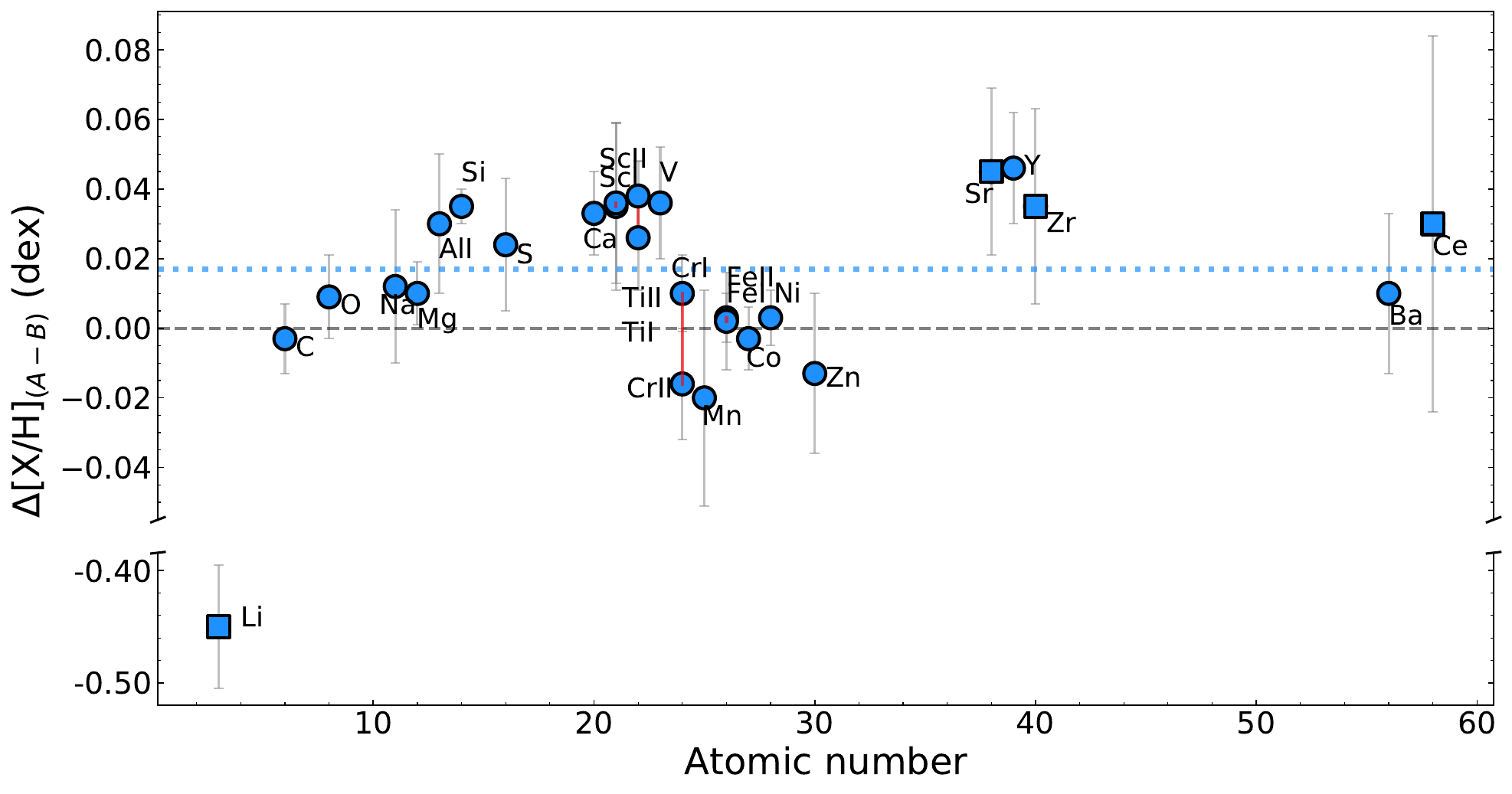}
\caption{Chemical composition difference between HD 202772A and HD 202772B as a function of atomic number. The dashed line corresponds to identical composition and the dotted line represents the weighted average of $\Delta$[X/H]$_{A-B}$ considering all the elements but those measured from only one line (see text for more details). Red vertical lines connect two species of the same chemical element (e.g., Sc \textsc{i} and Sc \textsc{ii}) and  squares show the species measured from only one line (Li \textsc{i}, Sr \textsc{i}, Zr \textsc{ii}, and Ce \textsc{ii}). \label{fig.abun.vs.Z}}
 
\end{figure*}

Figure \ref{fig.abun.vs.Z} shows the computed differential abundances of HD 202772A relative to HD 202772B, $\Delta$[X/H]$_{A-B}$, as a function of atomic number. We can see that HD 202772A is more metal-rich than its binary companion across several elements. Conservatively, if we exclude the abundances of those elements measured from only one line (i.e., Sr, Zr, Ce, and Li), we find a total weighted average difference (and weighted standard deviation) of $\Delta$[X/H]$_{A-B}$ = 0.016 $\pm$ 0.004 dex (i.e., an excess detected at the $\sim$4.5$\sigma$ level). Hereafter, we discuss our results excluding the group of elements measured from only one line\footnote{The results, however, remain almost unchanged if we include the elements measured from only one line.}. 

\begin{figure*}
\centering
\includegraphics[width=.7\textwidth]{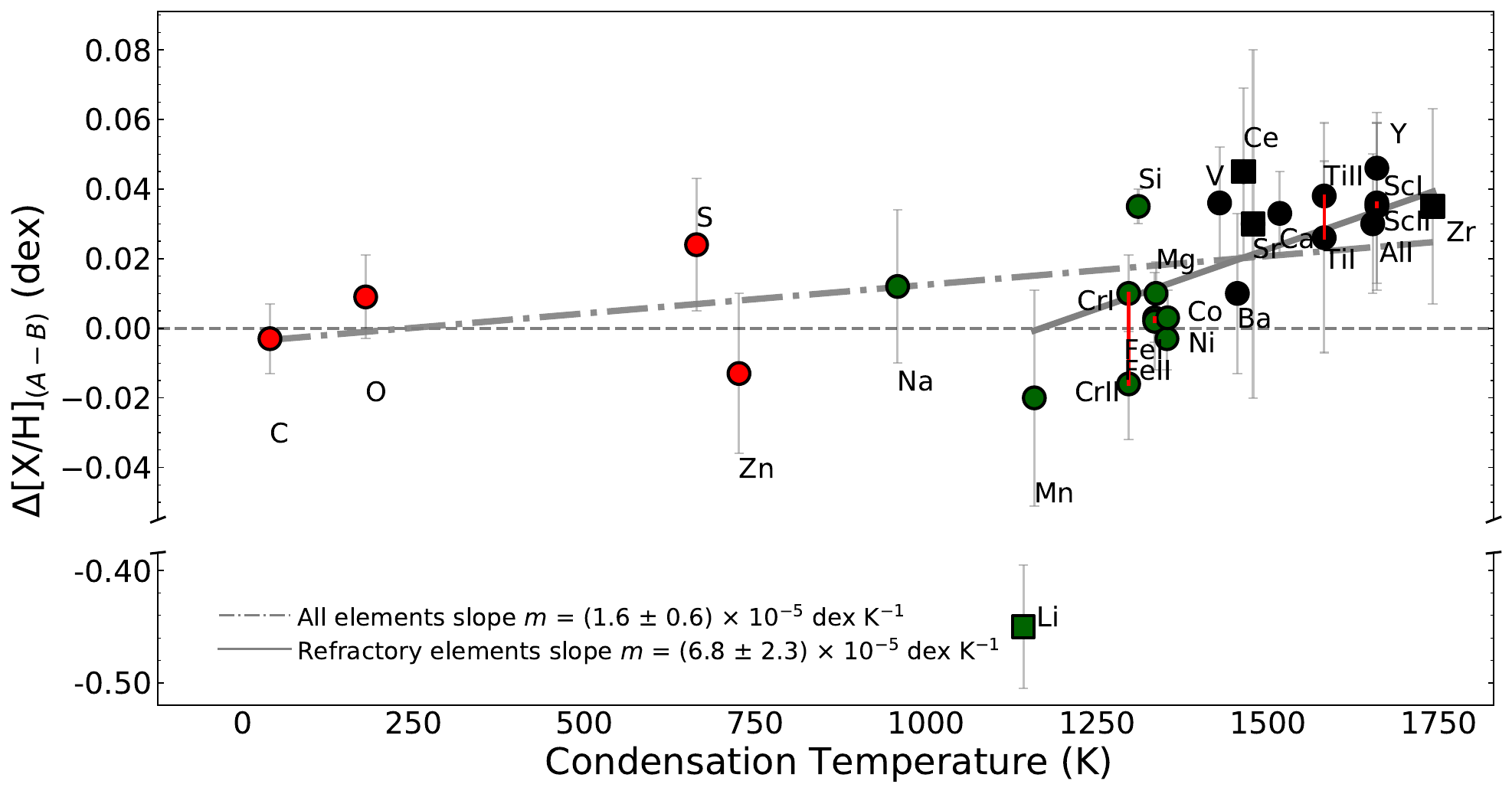}
\caption{ Chemical composition difference between HD 202772A and HD 202772B versus condensation temperature. Volatile elements (T$_{c}$ $\leq$ 1000 K) are shown in red, refractory elements with 1000 $<$ T$_{c}$ $\leq$ 1400 K are shown in green, while those with T$_{c}$ $>$ 1400 K are shown in black.
 The gray dot-dashed line is the weighted linear least-squares fit to all the elements measured from more than one line, whilst the solid one is for the refractory elements only (T$_{c}$ $>$ 1000 K). Squares show the species measured from only one line (Li, Ce, Sr, and Zr). As before, the black dashed line corresponds to identical composition.
\label{fig.Tcond.todo}}
\end{figure*}

In Figure \ref{fig.Tcond.todo}, we plot the differential abundances versus the 50\% condensation temperature (T$_{c}$) from \citet{Lodders2003} for solar-composition gas. Here, it can be seen that both stars show similar composition, within the errors, except for Si and other refractory elements with T$_{c}$ $>$ 1400 K. Taking into account all refractory elements (T$_{c}$ $>$ 1000 K), the average difference is $\Delta$[X/H]$_{A-B}$ = +0.018 $\pm$ 0.004 dex (i.e., an excess detected at the $\sim$4$\sigma$ level) and if we consider only those with T$_{c}$ $>$ 1400 K the average difference is $\Delta$[X/H]$_{A-B}$ = +0.035 $\pm$ 0.006 dex (i.e., an excess detected at the $\sim$6$\sigma$ level) relative to its companion without detected planets. 

 \begin{table}
      \caption{Tested models to check the correlation between the $\Delta$[X/H]$_{A-B}$ and T$_{c}$ data.}
         \label{table.models}
     \centering
         \begin{tabular}{l c c c}
            \hline
Model	&	$\chi_{r} ^{2}$	&	$\sigma_{f}$	&	BIC	\\
\hline
\multicolumn{4}{c}{Refractory elements only (measured from more than one line)} \\	
\hline
i- zero-slope and zero-intercept fit ($m$=$b$=0)	&	5.6	&	0.018	&	95	\\
i- weighted linear fit ($m$=0 and $b$)	&	2.73	&	0.017	&	46	\\
iii- weighted linear fit ($m$ and $b$ free)	&	2.37	&	0.012	&	41	\\
\hline
\noalign{\smallskip}
\end{tabular} 
\vskip 0.05 cm
\noindent {\footnotesize{Notes: -Reduced chi square: $\chi_{r} ^{2} = \chi^{2}/(n-d)$, where $n$ is the number of data-points,  $d$ is the number of free parameters in the model, $\chi^{2} = \Sigma [\Delta[X/H]-\Delta[X/H]_{mod}]^2/$Error$^2_{obs}$, and $\Delta[X/H]_{mod}$ represents the differential abundance predicted by the model. 

- $\sigma_{f}$: Residuals root mean square. 

-Bayesian Information  Criterion BIC$ = \chi^{2} + k_{f} \log N_{p}$ ; where $k_{f}$ is the number of free parameters in the model and $N_{p}$ is the number of data-points \citep{Schwarz1978}. }\\
}

        \end{table}   

To check and quantify a possible correlation between the $\Delta$[X/H]$_{A-B}$ and T$_{c}$, we started by applying a weighted linear fit to all elements following the procedure detailed in \citet{Jofre2021}. In addition to the elements measured from only one line, here we also exclude the differential abundance of lithium since it falls well outside of the general trend (see end of this section). This analysis yields a slope of $m$ = (1.5 $\pm$ 0.6) $\times$ 10$^{-5}$ dex K$^{-1}$ (i.e., a significance of the slope at the $\sim$2.5$\sigma$ level) and a Pearson’s correlation coefficient of $r$ $\sim$ 0.43. Therefore, as expected, a global T$_{c}$ trend is not significant.  

Following \citet{Jofre2021}, for the refractory elements, we tested three models: i) a zero-slope ($m$=0) and zero-intercept ($b$=0; i.e., no chemical difference) fit; ii) a weighted linear fit to the data with $m$=0 and free intercept (i.e., overall constant difference); and iii) a weighted\footnote{By 1/$\sigma^{2}_{\Delta[X/H]_{A-B}}$.} linear fit to the data with $m$ and $b$ as free parameters (i.e, T$_{c}$ trend). Examining the $\chi_{r} ^{2}$, $\sigma_{f}$, and BIC values obtained for each model, summarized in Table \ref{table.models}, we see that models ii (overall constant difference) and iii (T$_{c}$ trend) provide statistically better fits than the one where the A and B stars are chemically indistinguishable. In particular, the linear fit with unconstrained $m$ and $b$ supports a non-negligible slope, $m$ = (6.8 $\pm$ 2.3) $\times$ 10$^{-5}$ dex K$^{-1}$ (significance of $\sim$3$\sigma$ and a Pearson’s correlation coefficient of $r$ $\sim$ 0.75), indicating a tentative correlation.

However, the values presented in Table \ref{table.models} also indicate that the T$_{c}$ trend model does not provide a statistically significant improvement over the constant difference model.
Therefore, although we can rule out that stars A and B share the same chemical pattern for refractory elements, we cannot determine whether the observed differences are better explained by a constant metallicity offset or by a positive correlation with T$_{c}$. 




\begin{figure}
\centering
\includegraphics[width=.45\textwidth]{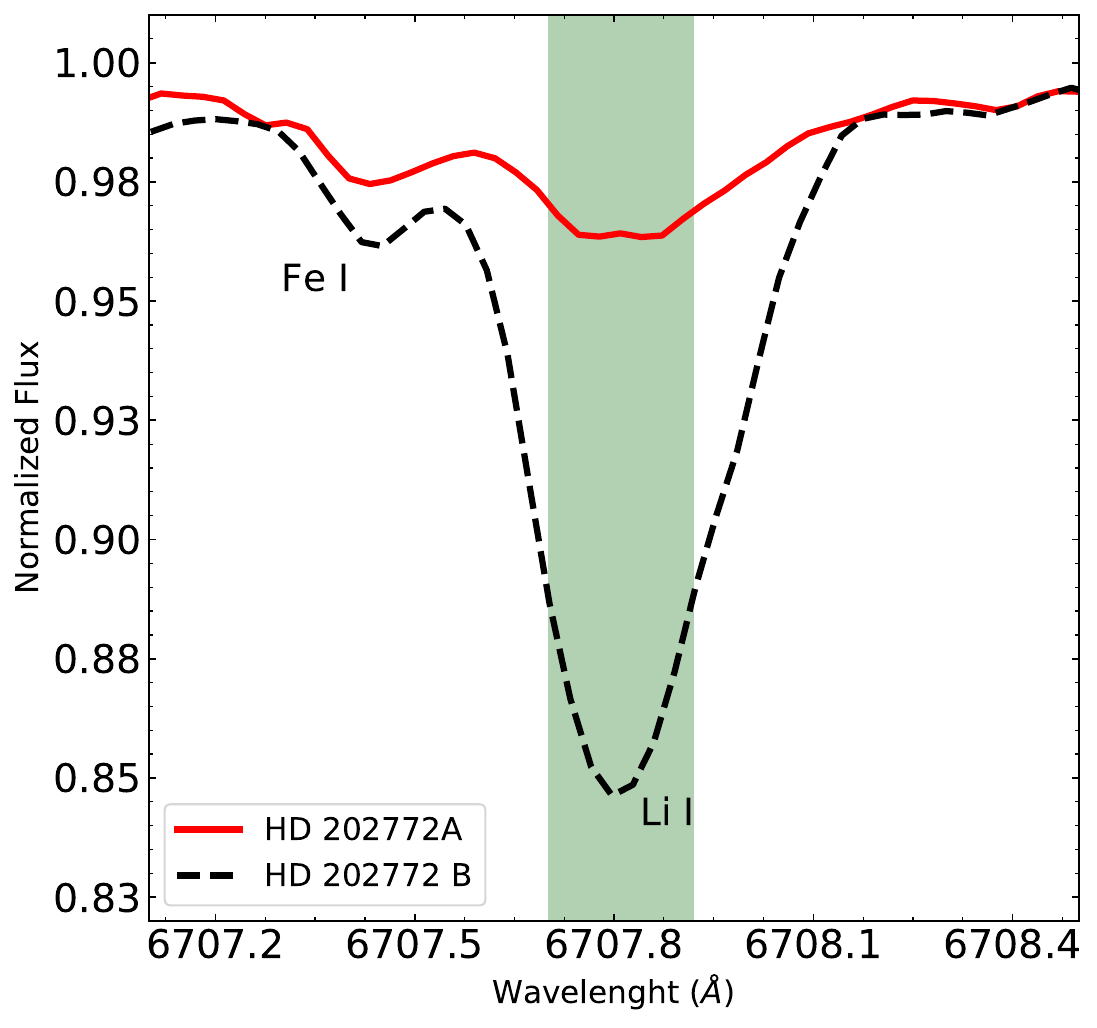}
\caption{Comparison between the observed spectra of HD 202772A (red line) and HD 202772B (black dotted line) around the 6607.8 {\AA} lithium line (shaded region). \label{fig.lithium.difference}}
\end{figure}

Finally, as can be seen in Figures \ref{fig.abun.vs.Z} and \ref{fig.Tcond.todo} 
the lithium differential abundance deviates from the general trend. In particular, the lithium abundance in the planet host star is 0.45 dex lower than the one measured on the companion star without detected planets, which can be clearly noticed in the observed spectra of both stars around the 6607.8 {\AA} lithium line (Figure \ref{fig.lithium.difference}). In Sec. \ref{sec.planet.engulfment} we discuss a possible scenario that might explain the observed difference.

\section{Planetary Analysis}\label{sec.planetary}

\begin{figure*}
\centering
\includegraphics[width=0.33\textwidth]{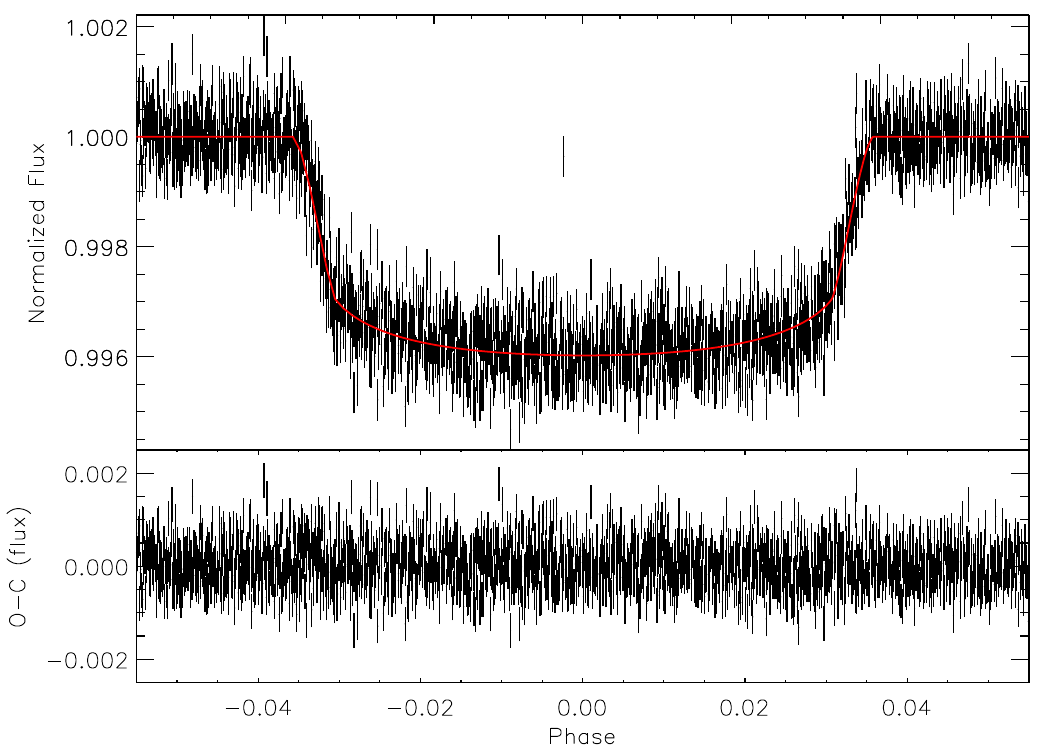}
\includegraphics[width=0.33\textwidth]{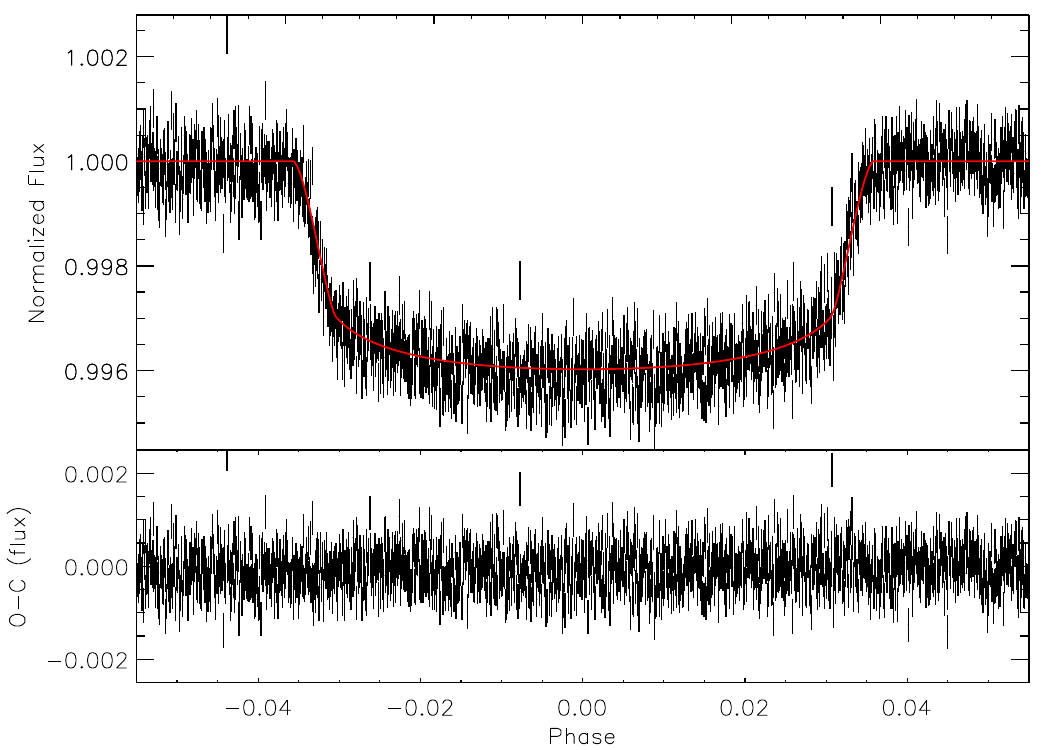}
\includegraphics[width=0.33\textwidth]{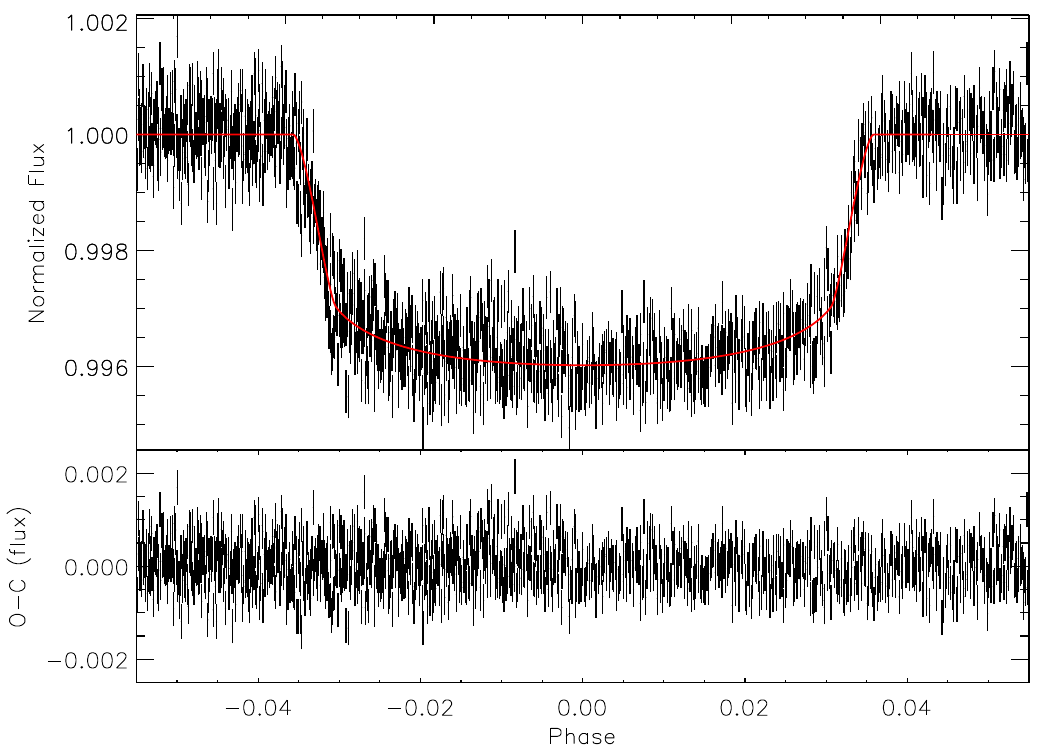}
\caption{Phase folded PDCSAP TESS transit light curves of HD~202772A~b. The observed data is shown by the black lines, the size of the line corresponds to the photometric uncertainty. The red solid line is the best-fit model from \texttt{EXOFASTv2}. The residuals are plotted below. The \textit{left} panel corresponds to Sector 1, the \textit{middle} panel to Sector 28, and the \textit{right} panel to Sector 68. 
\label{fig.tessLC}}
\end{figure*}

\begin{figure}
\centering
\includegraphics[width=.45\textwidth]{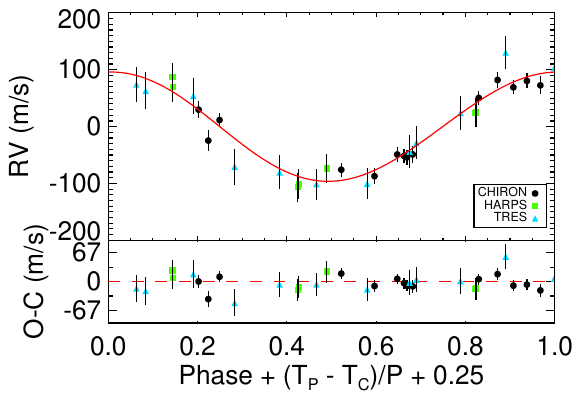}
\includegraphics[width=.44\textwidth]{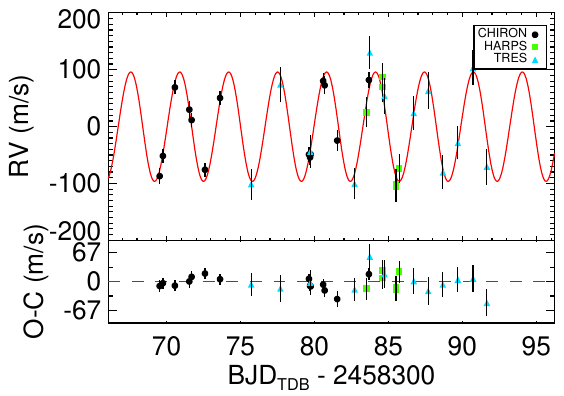}
\caption{{\it Top panel: } Radial velocity measurements of HD~202772A b from HARPS (green squares), CHIRON (black dots) and TRES (blue triangles) phase folded to the best determined period from our global model. The red solid line is the best-fit model from \texttt{EXOFASTv2}. The residuals are plotted below. Here $T_{P}$ is the time of periastron, $T_{C}$ is the time of conjunction or transit, and $P$ is the orbital period. 
{\it Bottom panel: } Radial velocity measurements as a function of time. There is no significant signal in the residuals (see also Fig.~\ref{fig.GLSomc}).} \label{fig.RV}
\end{figure}

\subsection{Refined planet properties of HD~202772A~b}{\label{exofastv2}}

We refined the bulk stellar and the planetary properties of HD~202772A and HD~202772A~b with the \texttt{EXOFASTv2} code \citep{Eastman2013, Eastman2019}, by simultaneously fitting the available PDCSAP TESS 2-min transit light curves (from sectors 1, 28 and 68\footnote{Some photometric data at the end of this sector were not considered due to the presence of systematics.}), the RV measurements from W19,  
literature broadband photometry including the photometry 
in the discovery paper, 
\textit{Gaia} DR3 parallax for HD~202772A, 
and the effective temperature and metallicity derived in Section~\ref{sec.atmospheric.parameters}. 
 The physical properties of the planetary system are derived by simultaneously fitting, through a differential evolution Markov Chain Monte Carlo, 
 all the available data described above to a keplerian orbital motion, the transit model \citep{mandelagol2002},
 a single-star SED, and the set of MIST stellar models \citep{Choi2016}. 

We set Gaussian priors for the stellar parameters with independent constraints, namely  T$_{\mathrm{eff}}$, [Fe/H], and parallax, with widths corresponding to their 1$\sigma$ uncertainties. We used a uniform prior and an upper limit for the extinction of A$_{\rm V} < $ 0.17236 mag to avoid unrealistic solutions based on galactic dust maps \citep[][W19]{Schlafly2011}. 
 We used as starting points in 
 M$_{\mathrm{\star}}$, and  R$_{\mathrm{\star}}$, the values derived with \texttt{q$^{2}$} presented in Section~\ref{sec.other.parameters}. Since these parameters are not independent from the \texttt{EXOFASTv2} analysis, we use Gaussian priors in the fitting with a width of 5$\times$ the 1-$\sigma$ uncertainty from the previous analysis. We note that we did not set any hard limit on the derived stellar mass and radius. 
 We employed Gaussian priors using the values and 1-$\sigma$ uncertainties derived in W19, for the orbital period, the eccentricity, the angle of periastron, and the inclination of the orbit. In the case that the uncertainty was not symmetric, we used the largest absolute value as the width of the Gaussian prior.  
 For the TESS light curves, we accounted for the 2-min cadence of the observations by setting EXPTIME to 2 and NINTERP to 1. To account for the dilution in the transit light curves due to the wide stellar companion HD~202772B, we fit for the FITDILUTE parameter using a Gaussian prior set to 0.209$\pm$ 0.1 (conservatively using 5$\times$ the 1-$\sigma$ uncertainty derived in W19).  
 We also let the quadratic limb-darkening coefficients be fit freely to the TESS light curves. 
 
 The best-fit solution is attained by starting the MCMC chains several times until all the data sets were optimally fitted.  
 Each MCMC runs until the two convergence criteria for each parameter \citep[number of independent draws T$_z >$ 1000 and Gelman-Rubin statistic $<$ 1.01; as described in][]{Eastman2013} or the maximum number of steps in that run are reached. We integrated our \texttt{EXOFASTv2} solutions several times following the approach suggested by \citet{Eastman2019}. After the first run, using the  priors from the previous run. The final solution was attained when all parameters converged on at least one of the convergence criteria, and the solution was indistinguishable from the previous solution within the uncertainties, and the measures of goodness of the fit did not improve.

 The derived physical properties for the host star HD~202772A 
are presented in Table~\ref{table.all.stellar.parameters}, 
and for the planet HD~202772A~b are given in Table~\ref{table.planetary.parameters}.  
 We adopt the median values of the posterior distributions of each best-fit solution, and 68$\%$ confidence intervals as their 1$\sigma$ uncertainties.
 We do not find evidence of transit timing variations (TTVs) in the system as the resulting O-C values are all within 1$\sigma$ of a linear ephemeris. In this case, each transit was fitted individually and then compared with the resulting mid-transit time using the ephemeris derived from the global fit to obtain O-C.
 Our refined planetary parameters are consistent within the errors with those found by W19, but are more precise because of the refined stellar properties and the two additional TESS sectors. 
 In particular, one of the largest improvements was for the orbital period with two orders of magnitude smaller uncertainty than that reported in W19.  The planet radius is slightly larger than that derived in the discovery paper, but the uncertainty is of the same order of magnitude. The planet mass and RV semi-amplitude are consistent in both value and uncertainty to that of W19. The impact parameter has a lower uncertainty of about half the uncertainty reported in W19. The equilibrium temperature is about 20\% more precise than reported by W19. 
 In Fig.~\ref{fig.tessLC}, we show the resulting  best-fit models for the TESS transit light curves, and in Fig.~\ref{fig.RV} we show the best-fit model to the radial velocity measurements.


\subsection{Search for additional planets in the system} \label{sec.search.planets}

To better put in context our abundance trend results, we searched for additional planets in the HD 202772A/B system.

\subsubsection{Radial velocity data for HD~202772A}
The Generalised Lomb Scargle periodogram \citep[GLS;][]{zechmeisterkurster2009} was computed for the velocity model residuals. An offset between velocity instruments was determined at the same time as the amplitude of the sinusoidal model for each frequency. The result is presented in Fig.~\ref{fig.GLSomc}. The evaluation of the peak significance was estimated by shuffling the velocity data and their errors 1000 times and recording the largest peak power obtained in each realization. The empirical distribution was used to compute the 90\%, 99\%, and 99.9\% percentiles, shown as horizontal lines in the figure. None of the peaks has power above the 99\% percentile, and were therefore not deemed significant. We do not identify any additional planetary signals in the available RVs for HD~202772A. No RVs data are available for HD 202772B.

\subsubsection{TESS light curves}\label{search.tess}


The search for additional transiting planets around the binary was carried out on the residuals from sectors 1, 28, and 68 obtained after removing the best-fit model found in Section \ref{exofastv2} to the PDCSAP 2-min cadence TESS light curve. We discarded some photometric data at the end of sector 68 due to the presence of systematics. We ran the Transit Least Squares code \citep[TLS,][]{Hippke2019A} on the remaining PDCSAP TESS residuals considering the values of the quadratic limb-darkening coefficients extracted from the TESS Input Catalog \citep[TICv8.2,][]{stassun2019}. No transit signal suggesting the presence of a planetary companion was detected. \\

\begin{figure}
\centering
\includegraphics[width=.5\textwidth]{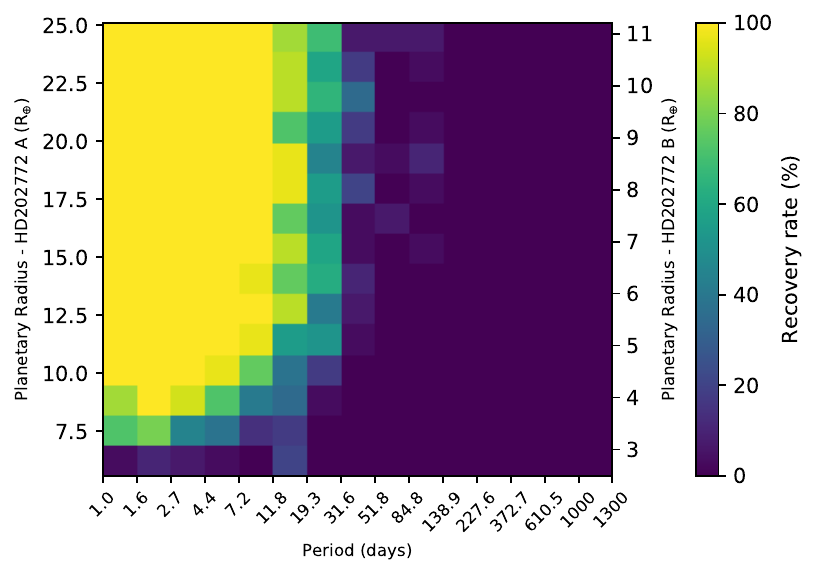}
\caption{Detectability map for HD~202772A and B performed on the PDCSAP TESS light-curve residuals. Light yellow-green and dark blue-violet colors indicate high and low recovery rates, respectively. Planetary radii on the right and left Y-axes are corrected by the contamination caused by the companion. 
\label{fig.injrec}}
\end{figure}

Furthermore, we performed an injection-recovery test on the PDCSAP TESS residuals of both stars in the system to assess the detection limits of additional transiting planets. We surveyed the planetary radius–orbital period parameter space, $R_{P}$-$P$, from $\sim$ 2.0 to $\sim$ 11.0 R$_{\oplus}$\footnote{R$_{\oplus}$ is the radius of Earth.} for the planetary radius and from 1 to 1300 days for the orbital period. For each $R_{P}$-$P$ combination, we adopted thirty different values of T$_{c}$, randomly selected between the minimum and maximum time-span values. The transit signals injected into the PDCSAP TESS residuals were created with the \texttt{BATMAN} code \citep{Kreidberg2015}, assuming equatorial transits (b $=$ 0), circular orbits (e $=$ 0), and a limb-darkening quadratic law. 
We considered a positive transiting planet detection when the recovered orbital period is within 5$\%$ of the injected period. The detectability map resulting from this test is shown in Figure \ref{fig.injrec}. Planetary radii on the right and left Y-axes were scaled by a factor that accounts for the contamination caused by the A and B companions, respectively. For the A component, this number was calculated through the variable FITDILUT, fit in \texttt{EXOFASTv2}. Meanwhile, for the B component, we used the dilution parameter ``D'' as stated in \citet{Jofre2021} to compute the same factor. As expected, it would be possible to find smaller transiting planets orbiting the MS star rather than the slightly evolved companion.

Additionally, Figure \ref{fig.injrec} shows that planets larger than $\sim$10 R$_{\oplus}$ and $\sim$ 4.5 R$_{\oplus}$ with orbital periods shorter than $\sim$12 days around HD~202772A and HD~202772B, might be detected with high probability ($>$ 80$\%$). Small planets with radii $\lesssim$ 7.0 R$_{\oplus}$ and radii $<$ 3.0 R$_{\oplus}$ for the A and B components, respectively, have almost no chance of detection. The same result is expected for planets with orbital periods longer than $\sim$31 days. For all the planets that fill the parameter space in between, the detection probabilities range from 20$\%$ to 80$\%$.

\section{Discussion} \label{sec.discussion}
\subsection{Origin of the chemical differences}

\subsubsection{Planet engulfment} \label{sec.planet.engulfment}
The chemical difference observed in Figure \ref{fig.Tcond.todo}, where the planet host star A is enhanced in refractory elements relative to the B companion, could indicate the ingestion of rocky material. As suggested in the case of other binary systems \citep[e.g.,][]{Mack2014, Biazzo2015, Ramirez2015, Teske2015, Teske2016a, Saffe2017, Jofre2021, Spina2021, Flores2022, Miquelarena2024, YanaGalarza2024}, rocky bodies could have been pushed or dragged onto HD 202772A during the inward migration of the hot Jupiter detected around this star. 

First, to assess the dynamical history of HD 202772A and its Hot Jupiter, we compared key characteristic timescales with the system's estimated age. We focus on three mechanisms: tidal circularization, von-Zeipel-Lidov-Kozai (vZLK) cycles, and general relativistic (GR) apsidal precession. The planet's measured eccentricity of $0.03^{+0.024}_{-0.019}$ is consistent with zero at the $1.4\sigma$ level, suggesting a nearly circular orbit. Although this does not confirm past high eccentricity, it does not rule it out either. If the planet previously had a more eccentric orbit, tidal dissipation within the planet could have damped it over time. Assuming a Love number $k_2=0.3$ and a dissipation factor $Q’=10^4$–$10^6$ as in \citet{YanaGalarza2024}, and using the formalism of \citet{Rice2022}, we estimate the circularization timescale to be between 1 and 100 Myr, which is significantly shorter than the system's age of $\sim$1.6 Gyr. This makes tidal circularization a viable mechanism for shaping the current orbit of the planet.

Given the hierarchical architecture of the system, vZLK cycles may also have influenced the evolution of the planet, periodically changing its eccentricity. Using the approach of \citet{Naoz2016} and \citet{Rice2023}, we estimate the vZLK timescale to be 10-100 Myr, depending on the range considered for the binary eccentricity ($e_B=0$–$0.9$) and a separation of 200 au. These timescales are also shorter than the system's age, suggesting that vZLK migration may have contributed to the inward evolution of HD 202772A b. If the planet initially formed farther out, the vZLK timescale would have been even shorter. However, vZLK cycles can be suppressed if other effects, such as GR-induced apsidal precession, act on shorter timescales. Using the formulation from \citet{Rice2023}, we estimate the GR precession timescale for HD 202772Ab to be $\sim10^4$ years, which is orders of magnitude shorter than the vZLK timescale. This rapid precession likely prevents vZLK oscillations from operating at present. However, vZLK may have played a role earlier in the system’s evolution, before the planet migrated inward and GR precession became dominant.

The timescales mentioned above suggest that the hot Jupiter orbiting HD 202772A may have formed farther out with a higher eccentricity, reaching its current orbit through inward migration. Therefore, as recently proposed for other binary systems \citep[][]{Teske2015, Saffe2017, Oh2018, Jofre2021, Flores2024, YanaGalarza2024}, this inward movement could have destabilized any interior terrestrial planets, ultimately causing them to be engulfed by the star. 

In addition, the relatively young age of the system ($\sim$1.4--1.9 Gyr) may provide further support for the planet engulfment scenario. Recently, \citet{Behmard2023} suggested that observable signatures of ingestion events are rarely detected in systems that are several Gyr old and tend to disappear $\sim$2 Gyr after the engulfment event.

Based on the results above, we tested whether an engulfment scenario could account for the observed anomalous chemical pattern. Thus, we used the last version of the code \texttt{terra} \citep{Galarza2016} to estimate the amount of rocky material (mix of meteoritic and terrestrial abundance) that needs to be added to the convective zone of HD 202772A to reproduce the observed abundances. In stars with masses around 1.5 $M_\odot$, the outer convective zone is expected to be extremely shallow \citep[e.g.,][] {Pinsonneault2001}. By adopting a value of $M_{cz}$ = 0.001 $M_{\odot}$ ($\sim$ 0.06\% of the stellar mass) from the relation in \citet{Murray2001}, we find that the engulfment of $\sim$0.13 $M_{\oplus}$ could explain the observed T$_{c}$ trend in the refractory elements. However, in line with the results from Section \ref{sec.differences.abundances} (i.e., the T$_{c}$ trend model does not provide a statistically significant improvement over the constant difference model), the predicted and observed differential abundances exhibit a poor level of agreement ($\chi_{r}^{2} \sim 2.4$)\footnote{In comparison, other planet-hosting wide binaries that display enhancements in refractory elements and have been recently analyzed with \texttt{terra}—such as HD 196067/68 \citep{Flores2024} and TOI-1173 A/B \citep{YanaGalarza2024}—show very good agreement between the observed and predicted abundances.}. This result, along with the absence of Fe abundance differences, which is a crucial factor in the engulfment scenario \citep[e.g.,][]{Spina2021}, makes this explanation unlikely for HD 202772A/B. Furthermore, this is consistent with the findings of BE23 for this binary and with recent studies reporting low ($\sim$5–8\%) planetary engulfment rates in binary systems \citep[e.g.,][]{Liu2024}.

\begin{figure}
\centering
\includegraphics[width=.49\textwidth]{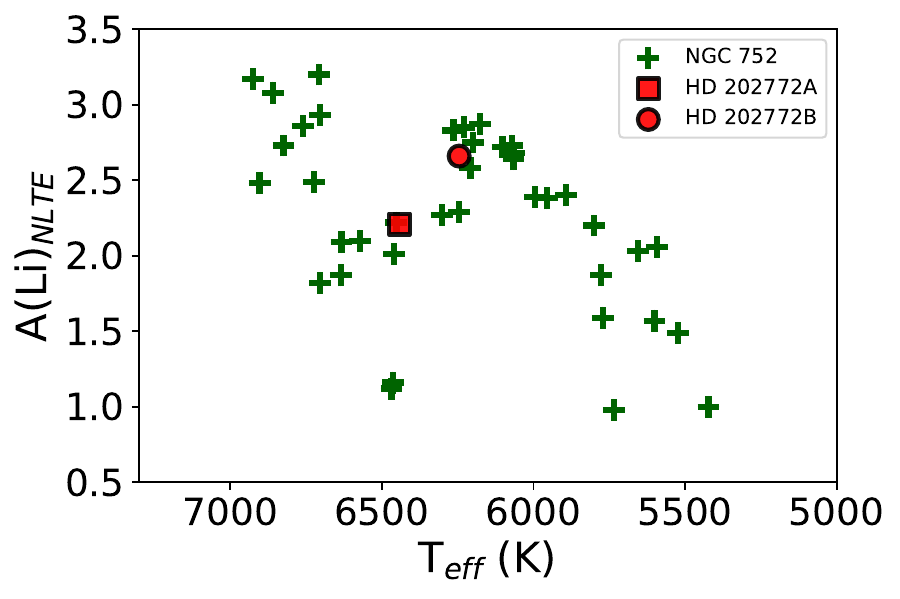}
\caption{A(Li)$_{NLTE}$ evolution vs. effective temperature for the open cluster NGC 752 and the binary system HD 202772A/B. The lithium abundances were taken from \citet{Boesgaard2022}. \label{fig.Li.dip}}
\end{figure}

Figure \ref{fig.lithium.difference} shows that the lithium line is weaker in the planet host star HD 202772A than in its companion without detected planets, HD 202772B. In particular, the lithium abundance in the A star is $\sim$0.45 dex lower than that measured on the B star. The observed lithium difference can be attributed to differences in the stellar parameters of the binary components \citep[][]{Sestito2005, Delgado2015, Ramirez2012, Castro2016, Bragaglia2018, Rathsam2023}. 

 The open cluster NGC 752 with [Fe/H] = +0.08 dex \citep[][]{Carrera2011} and age of $\sim$1.6 Gyr \citep[e.g.,][]{Castro2016}, can be used as a reference for the evolution of lithium \citep[Fig. 7 in][]{Boesgaard2022} in the HD 202772A/B system. As can be noticed in Fig. \ref{fig.Li.dip}, the observed difference in effective temperature (or mass) between the two stars can account for the 0.45 dex difference in lithium abundance. Notably, HD 202772B lies on the lithium plateau, where stars retain almost the maximum lithium content, while HD 202772A is located near the so-called lithium dip  \citep{Boesgaard1986}, where lithium depletion becomes more pronounced. 
A very similar approach was recently employed by \citet{Flores2024} to explain the lithium differences in the planet hosting binary HD196067-8. 


\subsubsection{Rocky or/and gas giant planet formation}
An alternative hypothesis proposes that the stellar surface could be depleted in refractory elements that were sequestered during the formation of rocky material (e.g., terrestrial planets, planetesimals) \citep[e.g.,][]{Melendez2009, Ramirez2009}. Under this scenario, the observed chemical pattern in Figure \ref{fig.Tcond.todo} could be interpreted as a signature of missing refractory elements in HD 202772B that were locked in rocky material. Again, we used the code \texttt{terra} to estimate the amount of refractory elements in rocky material locked around HD 202772B required to reproduce the observed differential abundances. 
We find that this amount is  $\sim$0.75 $M_{\oplus}$. Here, we adopted a  convective zone mass of $M_{\mathrm{cz}} \sim$ 0.005 $M_{\odot}$ based on the relation from \citet{Murray2001}\footnote{If we use the Yonsei–Yale isochrones \citep{Yi2001, Demarque2004} —the default models in the \texttt{terra} code, which are limited to stellar masses up to 1.3 $M_{\odot}$— we find that the mass of refractory elements locked in rocky material is $\sim$1.32 $M_{\oplus}$, assuming a convective zone mass of $M_{\mathrm{cz}}$ $\sim$ 0.009 $M_{\odot}$.}. 
However, as in the previous section, the poor level of agreement ($\chi_{r}^{2} \sim 2.4$) between the observed and predicted abundances makes it unlikely that the formation of a rocky planet could account for the observed chemical pattern.


On the other hand, recent studies, based on numerical simulations of protoplanetary discs, propose that the depletion of refractories can be caused by the gas–dust segregation process during the formation of gas giant planets \citep{Booth2020, Huhn2023}. In particular, the formation of a Jupiter-analog planet creates a gap— and pressure trap— that would prevent the accretion of the dust (refractories) exterior to its orbit onto the star in contrast to the gas (volatiles). 
Since these models currently do not predict a theoretical $T_{\mathrm{c}}$ trend suitable for direct comparison with the observations, a definitive interpretation remains challenging. Moreover, no planet has yet been detected around HD 202772B that could serve as detailed input for such models. Nonetheless, we cannot rule out that the refractory element deficit observed in the B component (relative to A) may have resulted from a gas–dust segregation mechanism during the early formation of a distant giant planet.

As detailed in Section \ref{sec.search.planets}, we searched for transiting planets around HD 202772B using the available TESS photometry, in an effort to investigate whether the presence of undetected planets could account for the observed chemical pattern. We found no transit signals and we were able to discard, with a probability higher than 80$\%$, the presence of close-in planets ($P$ $\lesssim$ 12 days) with radii larger than 10 R$_{\oplus}$ and 4.5 R$_{\oplus}$ around HD~202772A and HD~202772B, respectively (see Figure \ref{fig.injrec}). However, the analyzed TESS photometric data are inconclusive regarding the presence of small and/or longer-period planets ($P$ $\geq$ 31 days). Long-term RV follow-up and/or additional photometric monitoring might help to constrain the possible presence of planets around this star.


\begin{figure*}[h!]
\centering
\includegraphics[width=.4\textwidth]{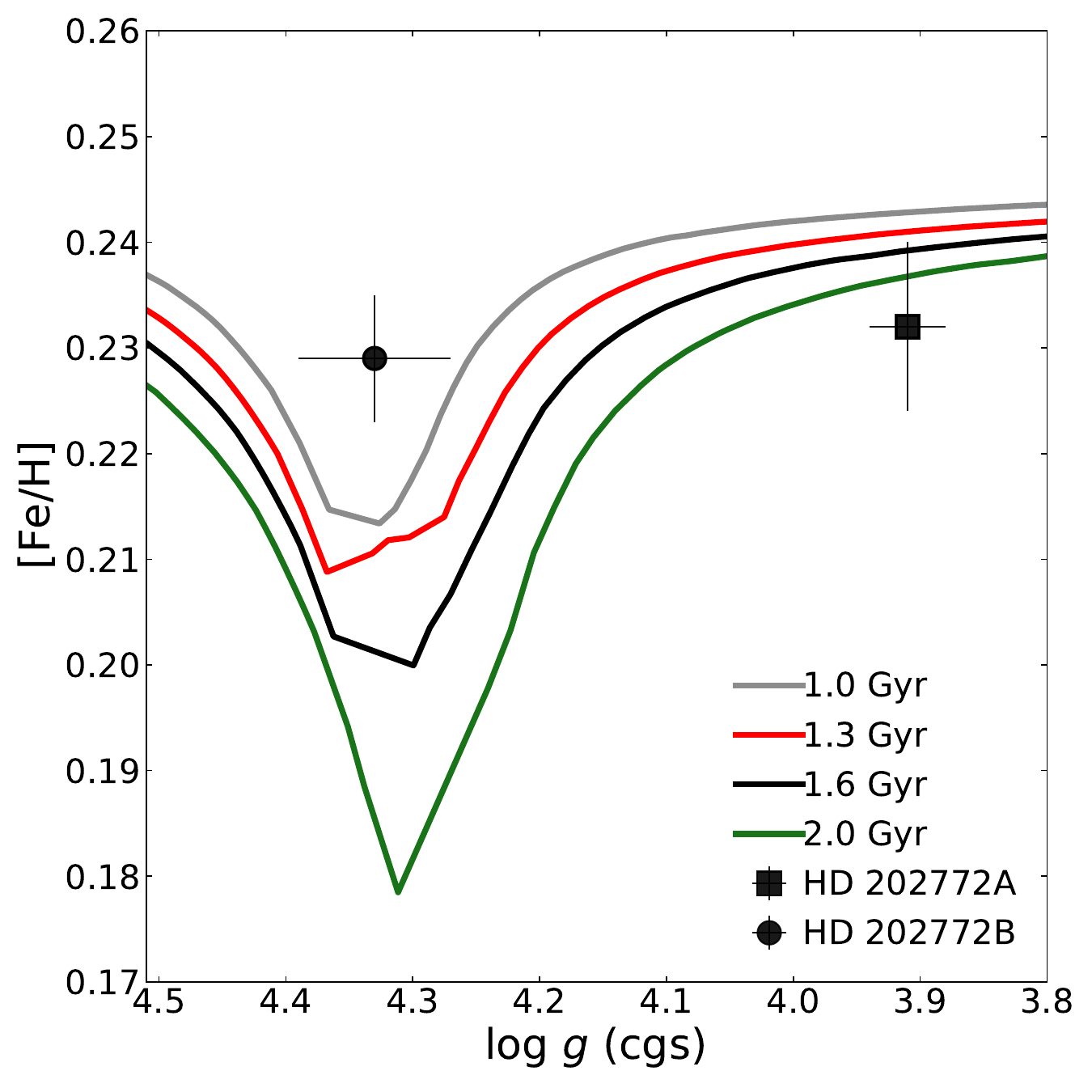}
\includegraphics[width=.4\textwidth]{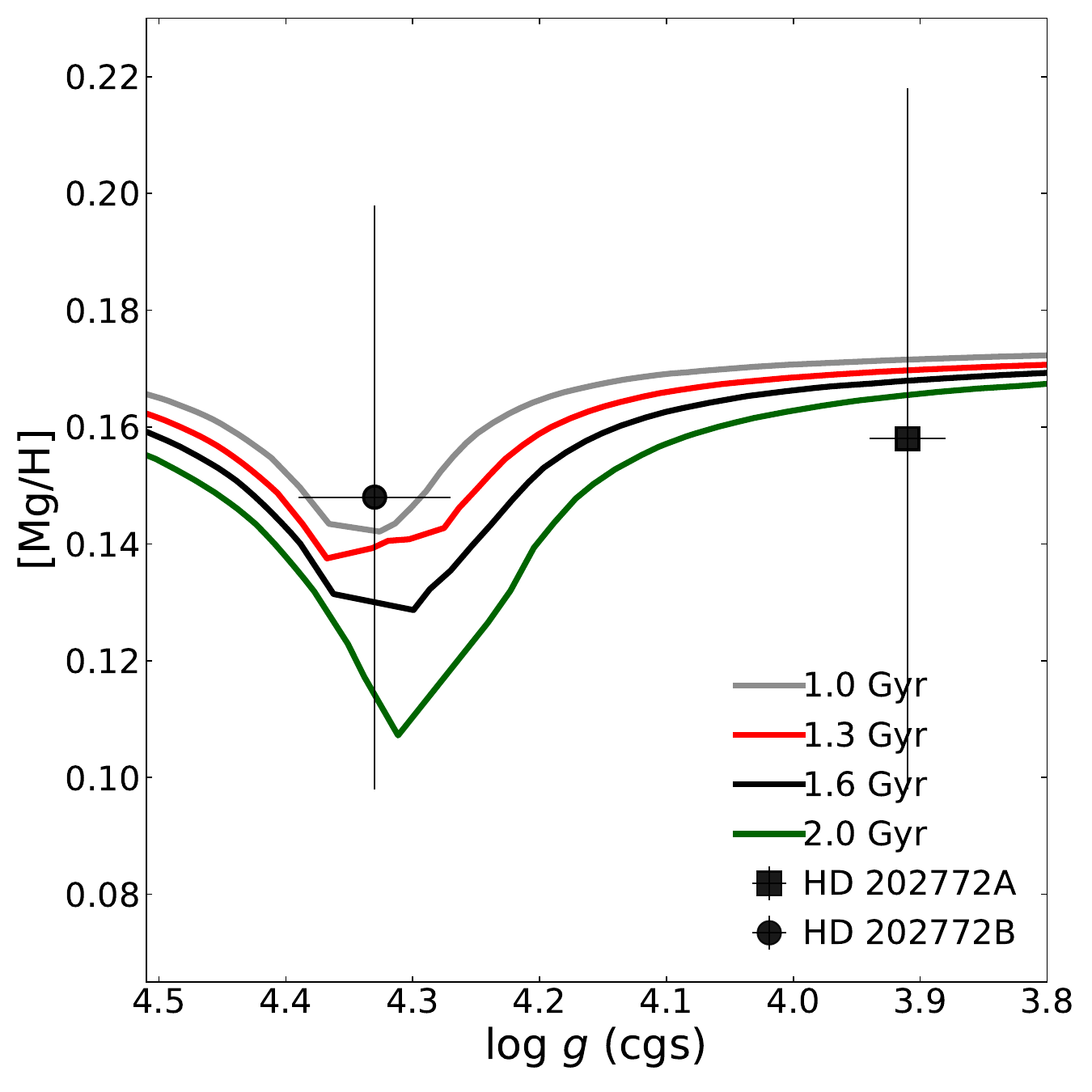}
\includegraphics[width=.4\textwidth]{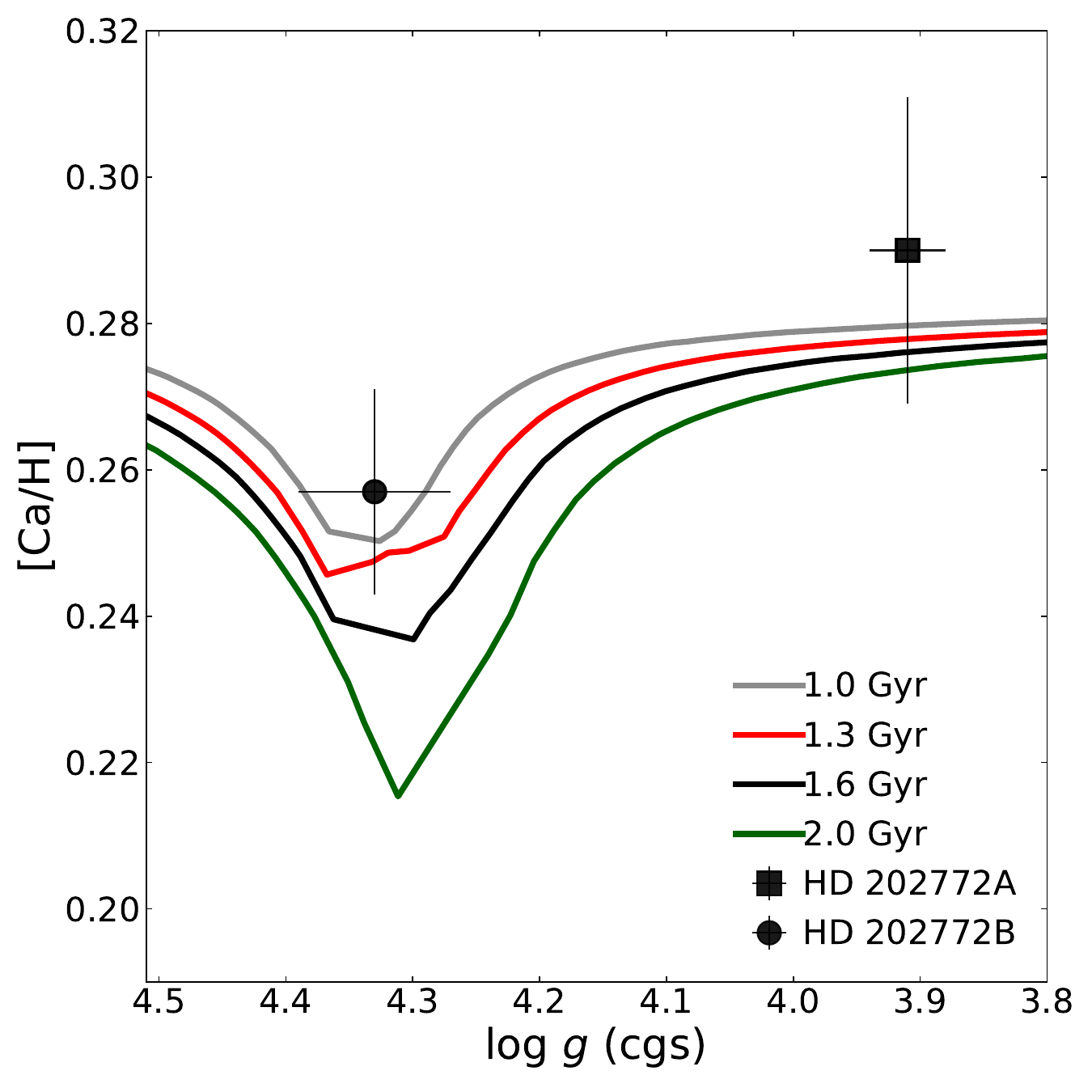}
\includegraphics[width=.4\textwidth]{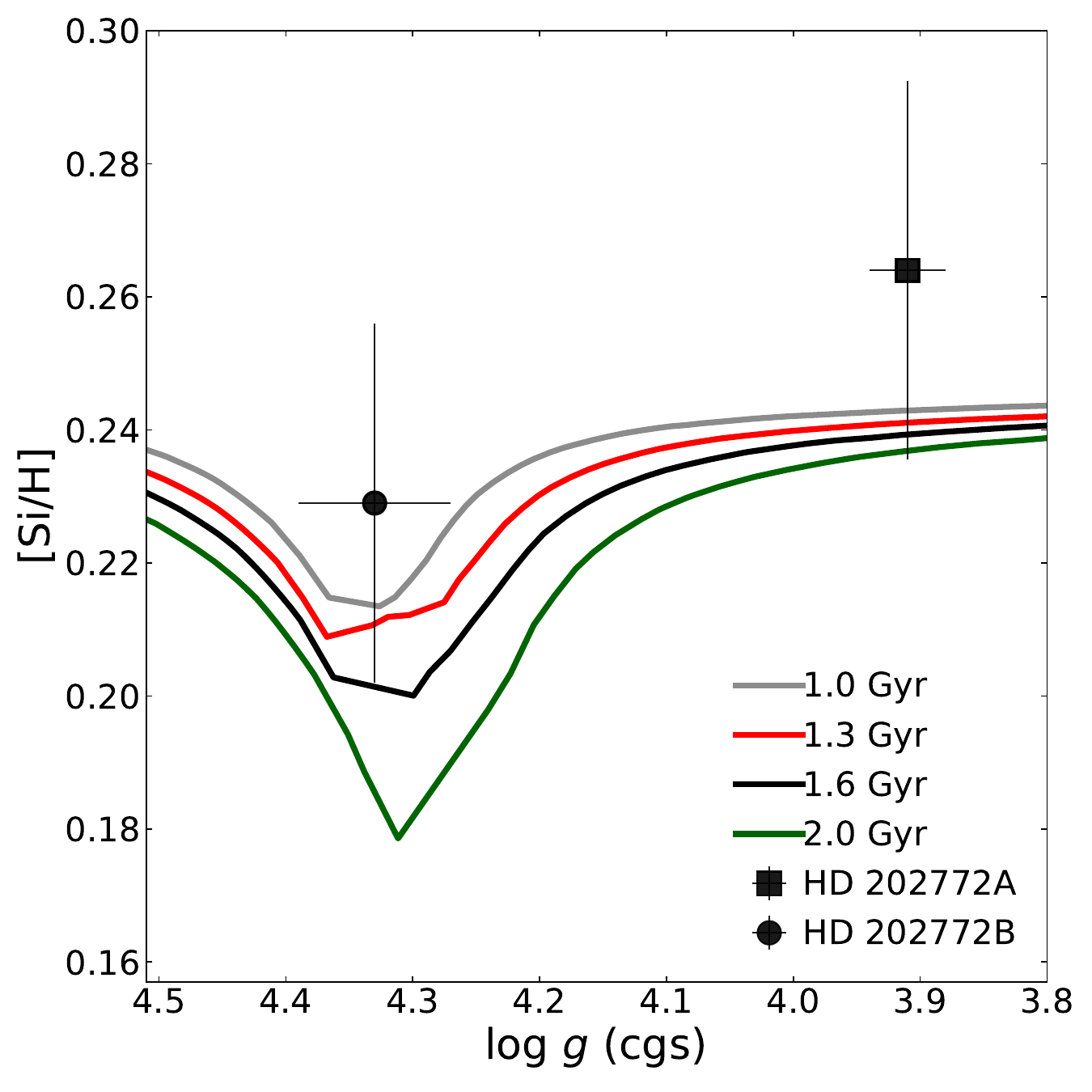}
\includegraphics[width=.4\textwidth]{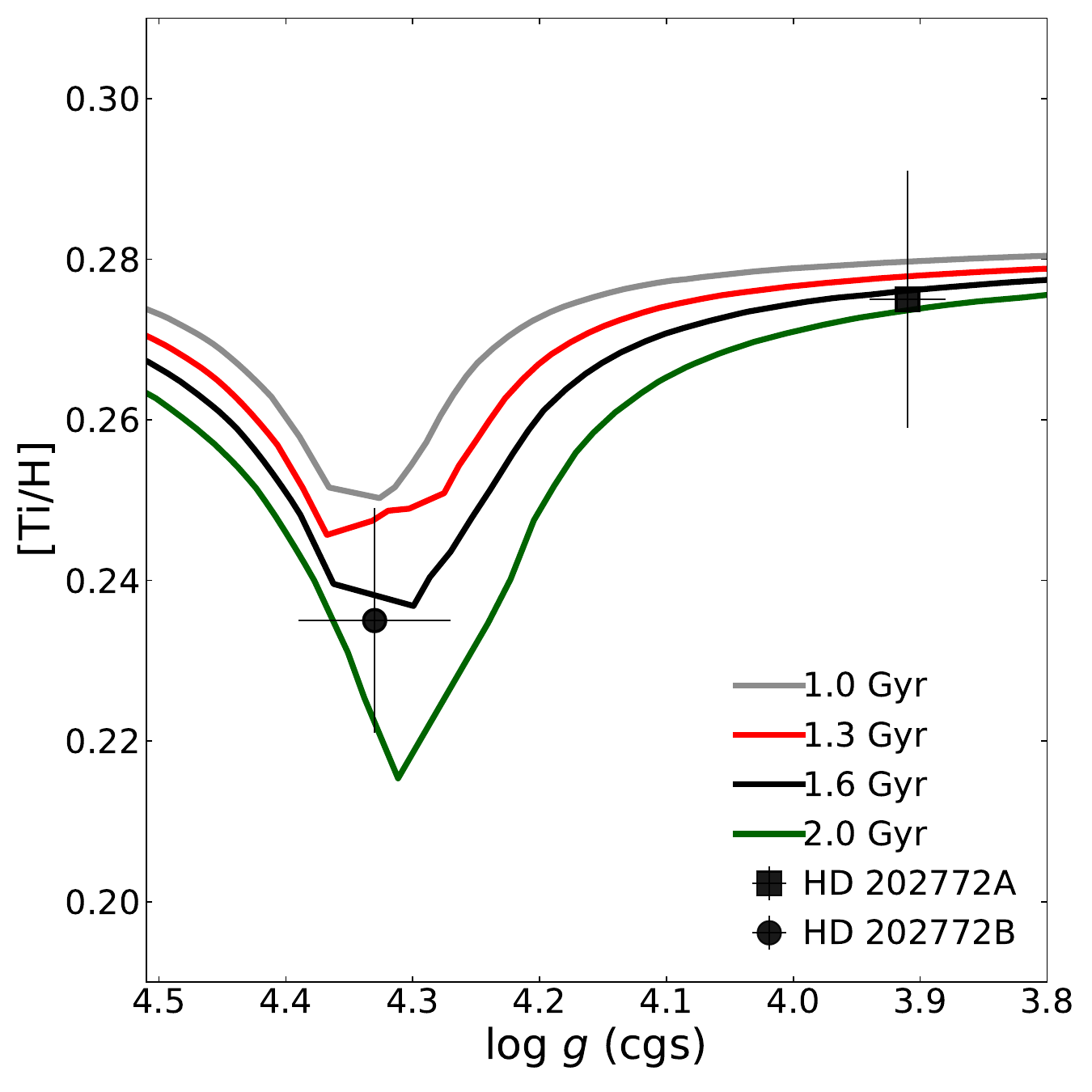}
\caption{Predicted abundances changes of Fe, Mg, Ca, Si, and Ti as a function of $\log g$, for ages between 1 and 2 Gyr due to atomic diffusion, as modeled by MIST \citep{Choi2016}. The squares and circles represent the measured abundances and $\log g$ values of HD 202772A and B, respectively. \label{fig.diffusion}}
\end{figure*}

\subsubsection{Effects of atomic diffusion}\label{sec.atomic.diffusion}
Another scenario that might account for the observed chemical differences is atomic diffusion. By combining the interplay of gravitational settling and radiative acceleration, atomic diffusion can alter the surface abundances of a star depending on its evolutionary stage \citep[e.g.][]{Dotter2017, Moedas2022}. Although the effects of atomic diffusion have been detected in stellar clusters for the past $\sim$20 years \citep[e.g.,][]{Korn2006, Souto2019, Liu2019}, only recently some possible signatures were reported in binaries \citep[e.g.,][LI21, hereafter]{Liu2021}. In particular, LI21 found that the overall abundance offsets between the binary components of 4 pairs could be attributed to the effects of atomic diffusion. 

Again, assuming that binaries form with the same initial composition, the chemical differences caused by atomic diffusion could be detected in pairs with relative large differences in the $T_{\mathrm{eff}}$  and $\log g$ (exceeding 200 K and 0.07 dex; LI21) of their components. 
Considering the difference in $T_{\mathrm{eff}}$  ($\sim$ 200 K) and especially in surface gravity ($\sim$0.4), which is the largest studied so far in MS wide planet hosting binaries, HD 202772A/B is an ideal system to search for signatures of atomic diffusion. 

We explored the atomic diffusion scenario using stellar evolutionary models for the average metallicity of the pair (0.23 dex)  and ages of 1.0, 1.3, 1.6, and 2 Gyr. Similar to \citet{Miquelarena2024} and \citet{YanaGalarza2024}, we employed MESA Isochrones and Stellar Tracks 
(MIST\footnote{\url{https://waps.cfa.harvard.edu/MIST/index.html}}) models \citep{Choi2016}, which include effects such as overshoot mixing and atomic diffusion. The results are presented in Figure \ref{fig.diffusion},
showing the predicted abundances for Fe, Mg, Si, Ca and Ti as a function of $\log g$ against the observed abundances of stars A and B (square and circle, respectively). Following MIST results, we should expect lower abundances for stars near $\log g \sim$4.3, that is, near the minimum of the v-shaped curves of Figure \ref{fig.diffusion}. In other words, MIST models predict lower abundances for star B ($\log g \sim$4.33) than for the more evolved star A ($\log g \sim$3.92).
The slightly lower abundances of Si, Ca, and Ti in star B compared to star A are, in principle, consistent with the expected diffusion models. 
However, considering the uncertainties in the abundances of Mg for both stars and the fact that the observed abundance of Fe for HD 202772B does not fall within the model predictions, the results should be interpreted with caution. Therefore, we conclude that although there may be signatures of atomic diffusion in the observed chemical pattern, this effect might not fully explain the chemical differences observed between HD 202772A and B.

\subsubsection{Other scenarios} 
$\delta$ Scuti stars are pulsating variables with $T_{\mathrm{eff}}$ values in a range of 6300 - 8500 K (i.e., $\sim$ A--F spectral types), $\log g$ values between 3.2 -- 4.3, and masses between 1.6 and 2.4 M$_{\odot}$ for near solar-metallicity stars \citep[e.g.,][]{Rodriguez2000, McNamara2011, Chang2013, Liakos2017}. The parameters derived for HD 202772A fall within the range of typical $\delta$ Scuti stars, close to the red edge of the instability strip in the HR diagram \citep[e.g.,][]{Liakos2017}. These variable stars pulsate in radial and non-radial modes with periods in the range $\sim$0.02--0.25 days (0.5--6 hours) with amplitudes between 0.003 and 0.9 mag in the V band \citep[e.g.,][]{Chang2013}. More importantly, this type of stars may be chemically peculiar. In particular, the prototype stars of this class ($\delta$ Scuti, $\rho$  Puppis) show excesses of up to 1.0 dex in the abundances of elements with Z $>$ 30 \citep{Yushchenko2005, Gopka2007}.

In an effort to explore the $\delta$ Scuti nature of HD 202772A,  we examined the 20-sec and 2-min cadence TESS photometry, ASAS, and the LCOGT data to search for hints of stellar pulsations. As mentioned in Sec. \ref{photvar}, we did not detect signs of variability. However, given that none of the observations used in this study, has the spatial resolution required to separate both components, HD 202772A's pulsations may still be diluted by the contamination flux of the B component. 

There are two $\delta$ Scuti stars that are more similar in stellar parameters to HD 202772A (in comparison to the prototype stars) with known detailed chemical abundances: CP Boo \citep{Galeev2012} and $\Psi^{1}$ Dra A \citep{Endl2016}. The abundances of elements with Z $>$ 30 in CP Boo are enhanced by about 0.3 dex. From Figure \ref{fig.abun.vs.Z}, we can see that the abundances of Zn (Z = 30),  Ba (Z = 56), and Ce (Z = 58) in HD 202772A agree, within the errors, with those in HD 202772B. The largest and most significant differences appear for Y (Z = 39, 0.046 $\pm$ 0.016 dex), Sr (Z = 38; 0.045 $\pm$	0.024 dex), and Zr (Z = 40, 0.035 $\pm$ 0.028 dex). Thus, considering the B star as reference and the large uncertainties in Ba and Ce, we could not discard the possibility that A is a $\delta$ Scuti star. However, when we analyze the measured abundances relative to the solar values, HD 202772A does not show an enhancement of heavy elements relative to light elements (Z $<$ 30). Here, as seen in CP Boo, a 0.3 dex enhancement in heavy elements would be easily detected in our data.  
Moreover, 3 out of 6 elements with Z $\ge$ 30 have been measured from one line only (Sr, Zr and Ce). Therefore, as we mention in Sec. \ref{sec.differences.abundances}, these elements were not included in the statistics (weighted average, linear fits, etc). To ensure that the observed trend is not driven by the other two elements (Y and Ba), which may be intrinsically enhanced in $\delta$ Scuti stars, we repeated the analysis from Section \ref{sec.differences.abundances}, excluding these two elements.  As the results remain largely unchanged, it is improbable that the chemical pattern in Figure \ref{fig.Tcond.todo} arises from an intrinsic chemical characteristic of HD 202772A linked to its potential $\delta$ Scuti nature.    

Interestingly, \citet{Endl2016} did not find an enhancement of elements with Z $\ge$ 30 for the other $\delta$ Scuti $\Psi^{1}$ Dra A which has very similar parameters to HD 202772A. This result would be in line with those of \citet{Fossati2008}, who do not find any substantial difference between the chemical pattern of $\delta$ Scuti stars and a sample of normal early A- and late F-type stars. Therefore, considering not only the chemical results but also the inconclusive photometric observations, we cannot rule out HD 202772A as a candidate for a star of the $\delta$ Scuti class.

Finally, in recent years, some studies have suggested that the chemical abundance differences observed in some binary systems could be the result of primordial inhomogeneities of the gas clouds from which the binary stars formed \citep[e.g.,][]{Ramirez2019, Liu2021, Behmard2023, Saffe2024}. The possible variation of the refractory content was also recently explored using numerical simulations of star formation \citep{Soliman2025}. However, \citet{Ramirez2019} suggest that the effect of chemical inhomogeneity might blur the T$_{c}$ trends observed in some binaries, but they do not compromise the significance of the T$_{c}$ correlation previously found on binary systems and/or their interpretation as due to sequestering or ingestion of planetary material. Furthermore, \citet{Ramirez2019} show that the primordial chemical differences appear to increase for larger separations between the stars in a binary system (see Fig. 8 and 9 from \citet{Ramirez2019}). 

More recently, in a pioneering work,  \citet{Saffe2024} found no clear T$_{c}$ trend but a significant overall abundance difference ($\sim$ 0.08 dex) between the components of a giant-giant binary system without known planets. If the metallicity offset was caused during the MS stage by a rocky planet locking or planetary engulfment, the difference should vanish as the stars evolve into red giants because their convective zones deepen and the metal-rich material dilutes. By ruling out recent planet-engulfment events and other scenarios, it was suggested that primordial inhomogeneities in the parent cloud might explain not only the observed differences in the giant-giant pair, but also in the other MS systems. A key conclusion by \citet{Saffe2024} is that binaries with overall abundance differences (volatile and refractory elements) but no clear T$_{c}$ trends are likely due to primordial inhomogeneities. 

Therefore, considering that: i) the projected separation between the components in HD 202772A/B  (200 au) is the smallest among the binaries with detailed abundance measurements, and ii) the chemical differences between the A and B components of HD 202772 do not show an overall constant offset, it is unlikely that  primordial inhomogeneities could explain the full observed abundance pattern.

\begin{figure*}
\centering
\includegraphics[width=0.8\textwidth,trim={2.1cm 0 4.1cm 1.3cm},clip]{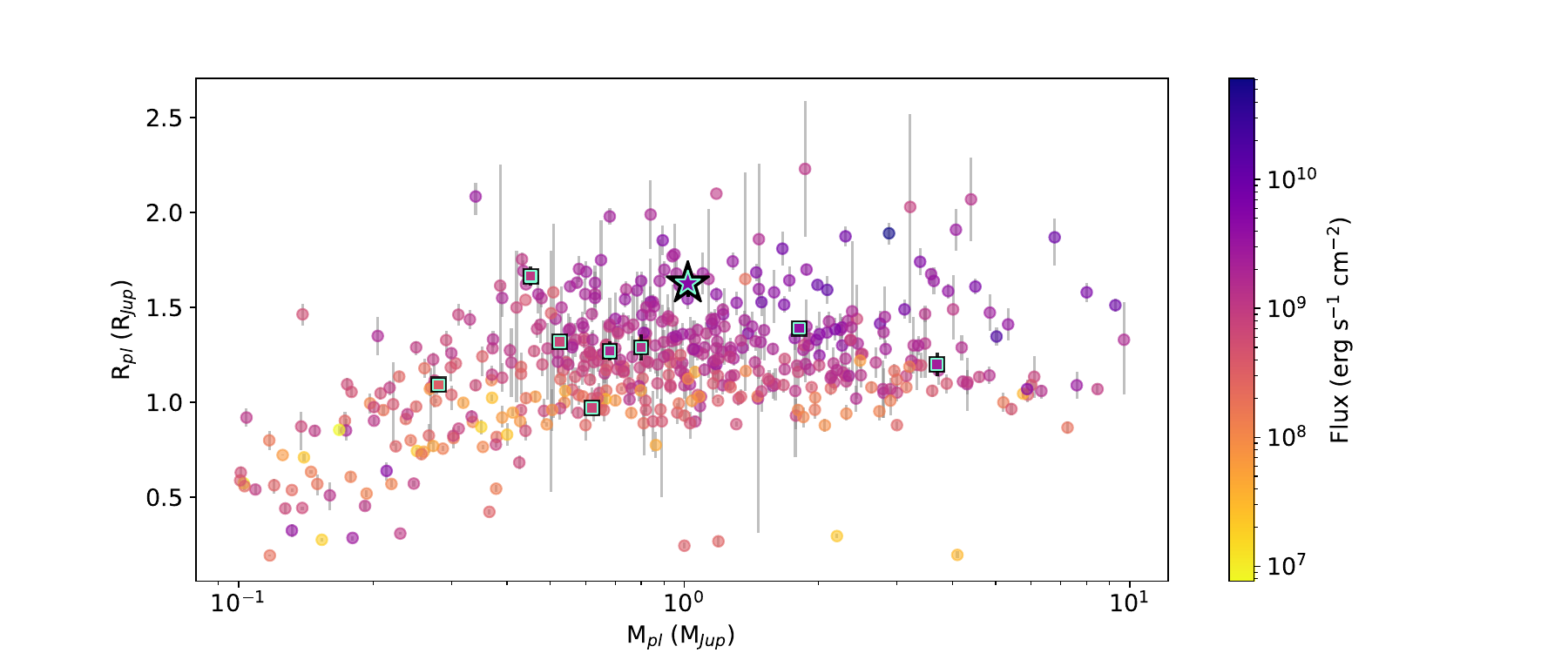}
\includegraphics[width=0.8\textwidth,trim={2.1cm 0 4.1cm 1.3cm},clip]{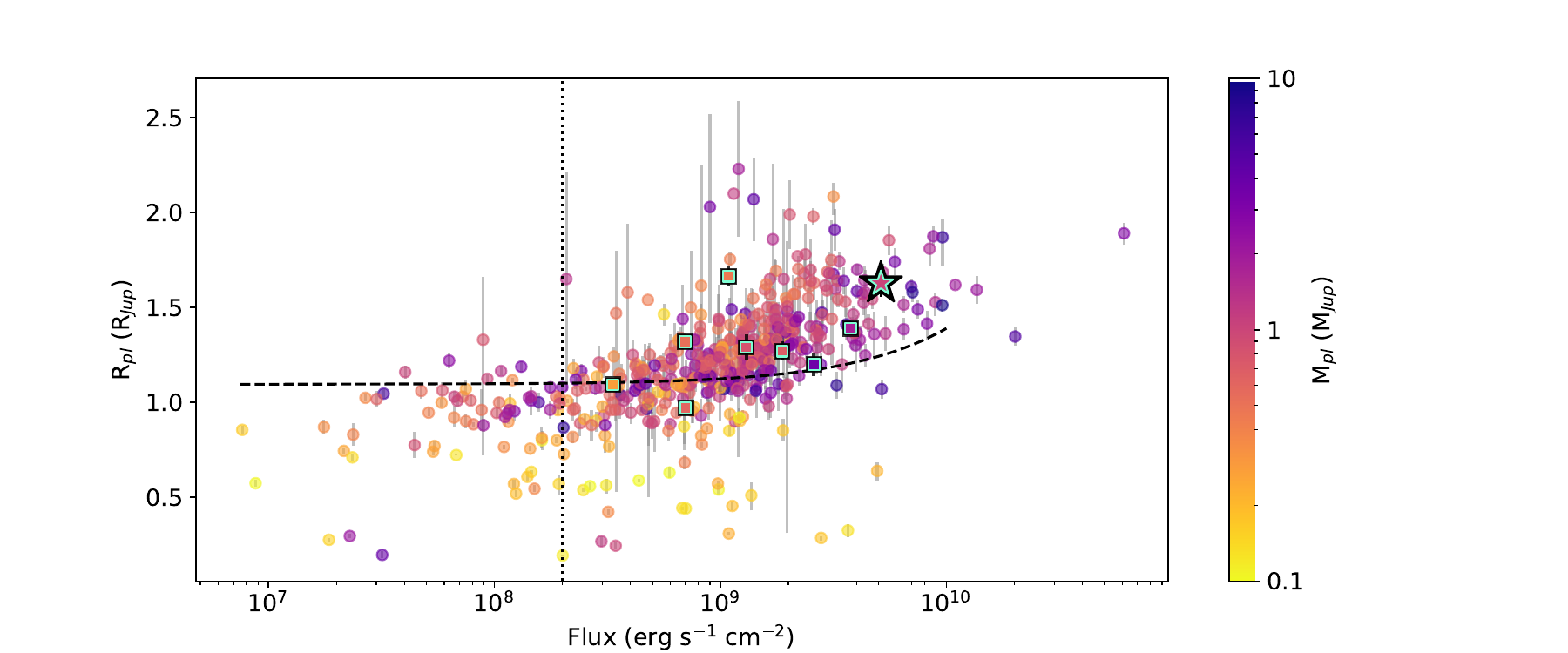}
\caption{HD~202772A b in the context of the known hot Jupiter population. 
    The best-fit parameters for HD~202772A b derived in this paper are shown by the black-and-cyan-edged star in both panels. The hot Jupiters orbiting one of the stars in a wide stellar binary for which precise properties are known are shown in the black-and-cyan-edged squares. Errors in the parameters are included and plotted in black for the wide-binary hot jupiters and HD~202772A~b. 
    The hot Jupiter sample (color circles) is selected from the NASA Exoplanet Archive  with errors in gray (retrieved on 2025-07-08). 
    {\it Top:} This planet mass--radius diagram shows that, in particular around 1~$M_{\rm Jup}$, the largest planets at a given mass are those that receive a higher total flux from its host star (in color bar). 
    {\it Bottom:} The radius vs incident flux also makes it clear. The vertical black, dotted line is at $2\times10^8$ erg~s$^{-1}$ cm$^{-2}$,  threshold above which inflation becomes significant \citep{Demory2011}.
        The black, dashed line is the 4.5 Gyr,  pure H/He, 1 M$_{\rm Jup}$ planet model with no added heating \citep[from Fig.~1 in][]{Thorngren2024}. 
    } \label{fig.hj}
\end{figure*}

\subsection{HD~202772A b in the context of the known hot Jupiters and upper limit on its remaining lifetime}

One of the remaining open questions regarding hot Jupiters is the nature of the radius anomaly or radius inflation \citep{Thorngren2024}. 
We compare HD~202772A~b with the known population of hot Jupiters in the literature, defined to have P$_{\rm orb} \le$ 10 days, 0.1 $\le M_{\rm pl} \le$ 10 $M_{\rm J}$, and 0.2 $\le R_{\rm pl} \le$ 2.5 $R_{\rm J}$ in the NASA Exoplanet Archive (retrieved on 2025-07-08). 
In this comparison we highlight the hot Jupiters orbiting stars in wide-binary systems namely, WASP-180A, HAT-P-1B, WASP-3A, WASP-94A, WASP-160B, HAT-P-4A, XO-2N and WASP-173A. In the case of WASP-180A and WASP-160B, we updated the parameters from papers that have precise stellar and planet parameters not reported in the archive \citep{Temple2019,Jofre2021}.
The hot Jupiters orbiting hosts in wide-binary systems have planet radii and planet masses that are within the ranges of the known hot Jupiters orbiting putative single stars, and are not more significantly inflated (see Fig.~\ref{fig.hj}).  HD~202772A~b is the hot Jupiter in the wide-binary sample receiving the highest irradiation and has a similar radius to WASP-94A~b within the quoted uncertainties.

 Using the empirical mass-flux-radius relation from 
 \citet{Weiss2013} 
 for planets with M$_{\rm pl}$ > 150 M$_{\rm \oplus}$ ($\sim$0.47~M$_{\rm J}$), we obtain a R$_{\rm pl} \simeq$ 15.9 R$_{\rm \oplus}$   ($\sim$1.42~R$_{\rm J}$) which is $\sim$4$\sigma$ away from our estimated planetary radius of 1.627 R$_{\rm Jup}$. This supports the correlation between radius inflation and stellar flux and strengthens the fact that the incident flux is a much better predictor of the radius anomaly than the period (see also Fig.~\ref{fig.hj}). Also, in this work we find the host star to be more evolved than previously thought, which could have an impact on the irradiated planet and its radius inflation \citep{Thorngren2019}. 
 Furthermore, inflated planets around evolved stars are key to better understanding the inflation mechanism \citep{Grunblatt2017}, 
 as these planets have been subjected to the stellar irradiation for longer timescales. For example HD~202772A~b could fall in the Class I from \citet{Lopez2016} 
 supporting the hypothesis that inflation requires direct heating.  


Finally, it is interesting to note that the planet-hosting star HD 202772A would be the typical progenitor of intermediate-mass evolved giant stars, for which there is an apparent scarcity of hot Jupiters \citep[e.g.,][]{Johnson2007, Niedzielski2009, Sato2008, Sato2010}. One of the main scenarios to explain the paucity of close-in planets considers that these objects might be engulfed by their host stars as they ascend on the red giant branch \citep[RGB; e.g.,][]{Villaver2009}. In this context, using our refined stellar parameters (listed on Table \ref{table.all.stellar.parameters}) and MIST stellar tracks \citep{Dotter2016}, we found that the stellar surface of HD 202772A will reach the orbit of the hot Jupiter planet in $\sim$0.55 Gyr ($\sim$11 R$_{\odot}$), almost halfway along the RGB. These values represent a conservative upper limit on the remaining lifetimes of the planet, because the engulfment would probably be accelerated due to tidal interactions in the star-planet systems, causing the orbital decay of the planetary companions \citep[e.g.,][]{Villaver2009, Matsumura2010, Kunitomo2011, Ronco2020}.

\section{Conclusions} \label{sec.conclusions}
We have conducted a detailed and precise spectroscopic and photometric characterization of the planet-hosting binary system HD 202772A/B. No planet has yet been reported around the B component, while the A star hosts a hot Jupiter transiting planet discovered from TESS data. For this planet, we also derived refined properties by performing a global fit that includes our precise stellar parameters, RV, and new sector TESS observations. We re-analyzed the available RVs data for HD 202772A and did not find any additional planetary signals. Furthermore, we used the TESS photometry to constrain the presence of transiting planets around HD~202772B and additional ones around HD~202772A. Our injection-recovery analysis allowed us to discard, with high probability ($>$80\%), the presence of transiting planets with orbital periods $\lesssim$12 days, and radii $\gtrsim$10 R$_{\oplus}$ and $\gtrsim$ 4.5 R$_{\oplus}$ around A and B, respectively. 


Another interesting property of the system is that  HD 202772A is about to leave the MS phase. Importantly, the irradiation received by the hot Jupiter is significantly higher than when the star was on the ZAMS, and it is the most irradiated of the known hot Jupiters orbiting a host in a wide binary. The refined planet mass and radius in this paper can be used to constrain the planetary heating efficiency, which may be as low as 0.03\% \citep{Grunblatt2017}, and which is key to understanding the radius anomaly of hot Jupiters. 

The spectroscopic characterization included, for the first time, strictly line-by-line differential fundamental parameters and precise chemical abundances for 27 species using high-quality Gemini-GRACES spectra.  We found that HD 202772B is 0.45 dex enhanced in Li I, relative to A, which can be explained by considering their difference in stellar temperature, mass and surface gravity. Moreover, the derived chemical pattern reveals that both stars have a similar content of volatile elements. However, the host star HD 202772A is slightly enhanced in refractory elements (especially for those with T$_{c}$ $>$ 1400 K), by +0.018 $\pm$ 0.004 dex, relative to the B component. We explored several scenarios to explain the observed differences.

One notable property of HD 202772A/B is the significant difference in the surface gravities of its components ($\Delta\log g_{A-B}\sim$  0.4 dex), which is one of the largest observed among the MS wide binaries analyzed to date. This allowed us to check, for the first time, the impact of atomic diffusion on the chemical pattern of this system. We found that the chemical differences for some refractory elements (especially Ti, Si, and Ca) could be attributed to this physical process. If confirmed, HD 202772A/B would be added to the very short list of planet-hosting wide binaries that exhibit possible signatures of atomic diffusion in their chemical patterns. 

We discussed two additional scenarios that are not related to planets. Specifically, we examined: i) the possibility that HD 202772A could be a chemically peculiar $\delta$ Scuti type star, and ii) the idea of a primordial origin for the chemical anomalies. However, we concluded that it is unlikely that either of these alternatives can fully account for the observed abundance pattern.

Among the planet-related scenarios, we explored the dynamical timescales of the giant planet detected around HD 202772A and found that the hot Jupiter might have arrived at its observed short period orbit through high-eccentricity migration (via tidal circularization and the von-Zeipel-Lidov-Kozai perturbation), as their timescales are significantly shorter than the age of the binary system. Thus, it is possible that the migration of the giant planet HD 202772A b could have triggered the fall of planetary material onto the surface of its host star. However, the poor agreement between the observed abundances and those predicted by the engulfment model, combined with the absence of significant Fe abundance differences, suggests that this scenario is unlikely to account for the observed chemical pattern.


Alternatively, although the formation of a rocky planet is also not supported by our analysis, we cannot rule out that a gap created by a long-period gas giant planet around HD 202772B might explain the chemical differences in refractory elements. However, so far, no evidence of planets has been reported around this star. Future long-term, high-precision photometric and RV follow-up, might provide further constraints on the real origin of the detected chemical differences.
 

Finally, although our data suggest  atomic diffusion as the most plausible scenario to explain the peculiar chemical differences observed in the HD 202772A/B system, we cannot rule out the contribution of additional processes. In a speculative scenario, the complex abundance pattern may result from a combination of processes, such as giant planet formation around HD 202772B coupled with the effects of atomic diffusion in HD 202772A. Notably, HD 202772A is not only the hottest component among the planet-hosting wide binaries analyzed to date, but also part of a class of systems that remain rare in current samples. Expanding the analysis to additional systems with similarly hot components is essential to further advance our understanding of the star–planet chemical connection.

\section*{Acknowledgements}

We thank the anonymous referee for a careful review and for the valuable comments and suggestions that helped us improve the manuscript.
We are grateful to F. Liu for providing valuable guidance on the use of MIST models.
The team acknowledges J. Eastman for his advice and support in using EXOFASTv2 and A. Behmard for generously providing the full data used in her study. 
This study is based on observations obtained through the Gemini Remote Access to CFHT ESPaDOnS Spectrograph (GRACES). ESPaDOnS is located at the Canada-France-Hawaii Telescope (CFHT), which is operated by the National Research Council of Canada, the Institut National des Sciences de l’Univers of the Centre National de la Recherche Scientifique of France, and the University of Hawai’i. ESPaDOnS is a collaborative project funded by France (CNRS, MENESR, OMP, LATT), Canada (NSERC), CFHT and ESA. ESPaDOnS was remotely controlled from the international Gemini Observatory, a program of NSF’s NOIRLab, which is managed by the Association of Universities for Research in Astronomy (AURA) under a cooperative agreement with the National Science Foundation on behalf of the Gemini partnership: the National Science Foundation (United States), the National Research Council (Canada), Agencia Nacional de Investigaci\'on y Desarrollo (Chile), Ministerio de Ciencia, Tecnolog\'ia e Innovaci\'on (Argentina), Minist\'{e}rio da Ci\^{e}ncia, Tecnologia e Inova\c{c}\~{a}o (Brazil), and Korea Astronomy and Space Science Institute (Republic of Korea). 

R.P. and E.J. acknowledge funding from CONICET, under projects PIBAA-CONICET ID-73811 and  ID-73669. This research has been partially supported by the Universidad Nacional Aut\'onoma de M\'exico (M\'exico) via PAPIIT IG-101224. MPR is partially supported by PICT-2021-I-INVI-00161 from ANPCyT, Argentina. J.Y.G. acknowledges support from a Carnegie Fellowship. E.M. acknowledges funding from FAPEMIG under project number APQ-02493-22 and a research productivity grant number 309829/2022-4 awarded by the CNPq.

Funding for the TESS mission is provided by NASA’s Science Mission Directorate. We acknowledge the use of public TESS data from pipelines at the TESS Science Office and at the TESS Science Processing Operations Center. Resources supporting this work were provided by the NASA High-End Computing (HEC) Program through the NASA Advanced Supercomputing (NAS) Division at Ames Research Center for the production of the SPOC data products.

This work makes use of observations from the Las Cumbres Observatory global telescope network. This research has made use of the NASA Exoplanet Archive, which is operated by the California Institute of Technology, under contract with the National Aeronautics and Space Administration under the Exoplanet Exploration Program.


\section*{Data Availability}
The TESS data are accessible via the Mikulski Archive for Space Telescopes (MAST) portal at https://mast.stsci.edu/portal/Mashup
/Clients/Mast/Portal.html.

The spectroscopic data are available from the corresponding author upon request.



\bibliographystyle{mnras}
\bibliography{example-jofre} 



\appendix

\section{Additional Figures and Tables}


\begin{figure*}
\centering
\includegraphics[width=.33\textwidth]{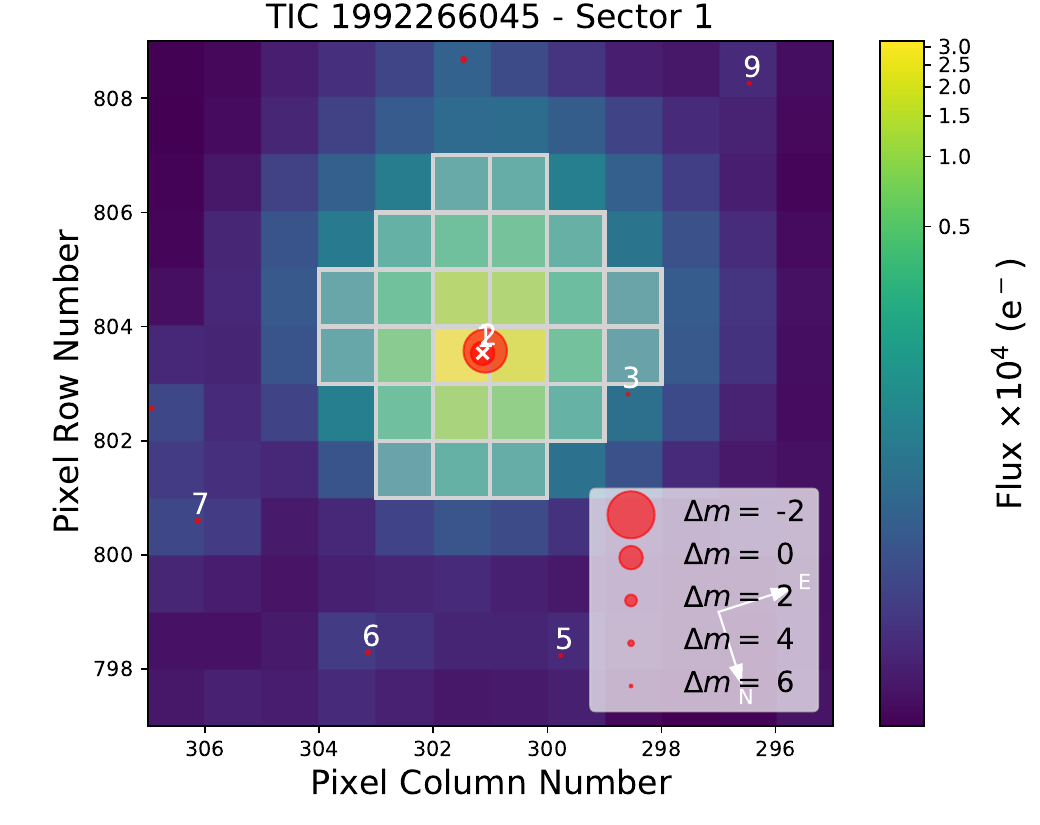}
\includegraphics[width=.33\textwidth]{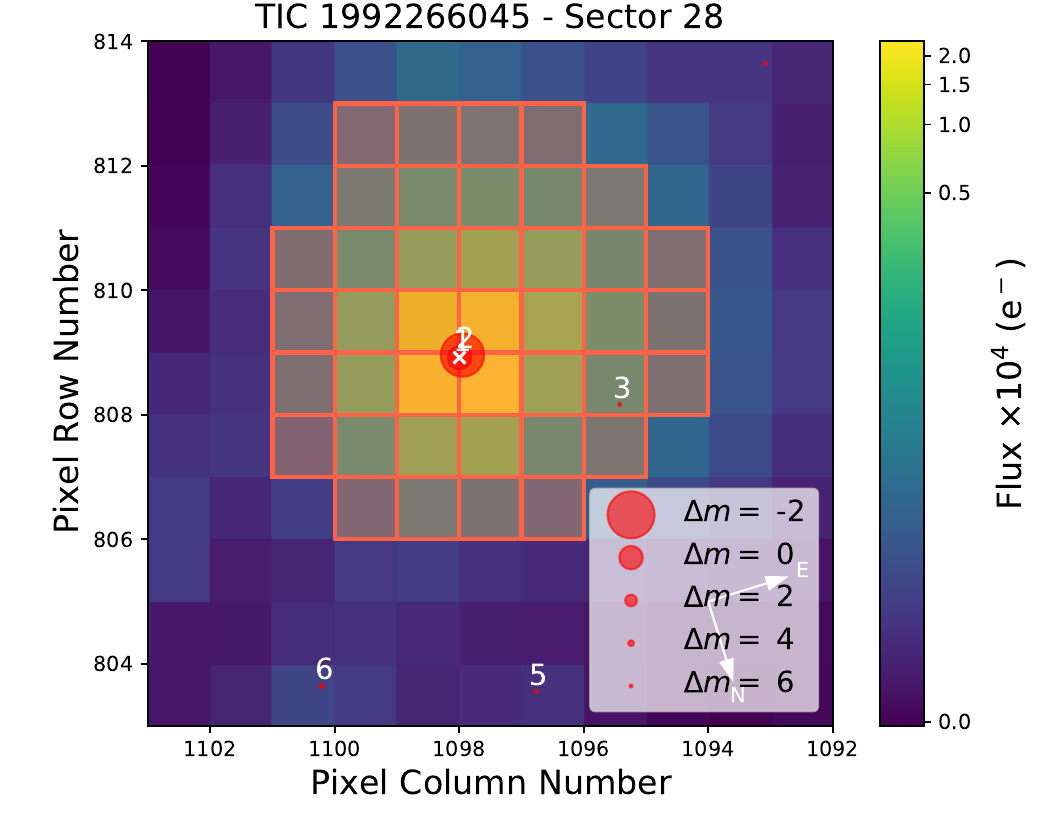}
\includegraphics[width=.33\textwidth]{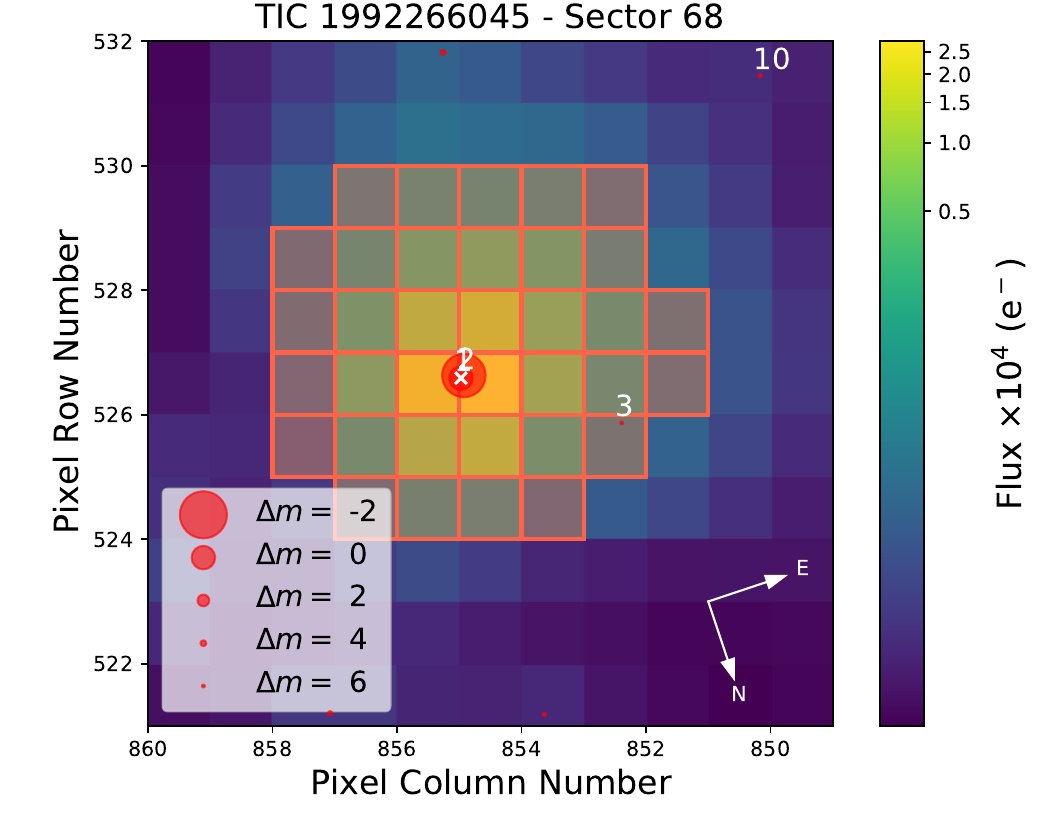}
\caption{Target pixel files of HD~202772A/B observed in TESS sectors 1 (\textit{left}), 28 (\textit{middle}), and 68 (\textit{right}). White crosses labeled as 1 and 2 point out the A and B components of the binary pair. In all the panels, both stars fall in the same pixel preventing them from resolving. Red circles are Gaia DR3 sources and the symbol sizes indicate the brightness of each source compared to that of the primary of the system. Shaded squares specify the pixels used in aperture photometry. The three plots were created with the TPFPLOTTER tool \citep{Aller2020}.\label{fig.TPF}}
\end{figure*}

\begin{figure}
\centering
\includegraphics[width=.45\textwidth]{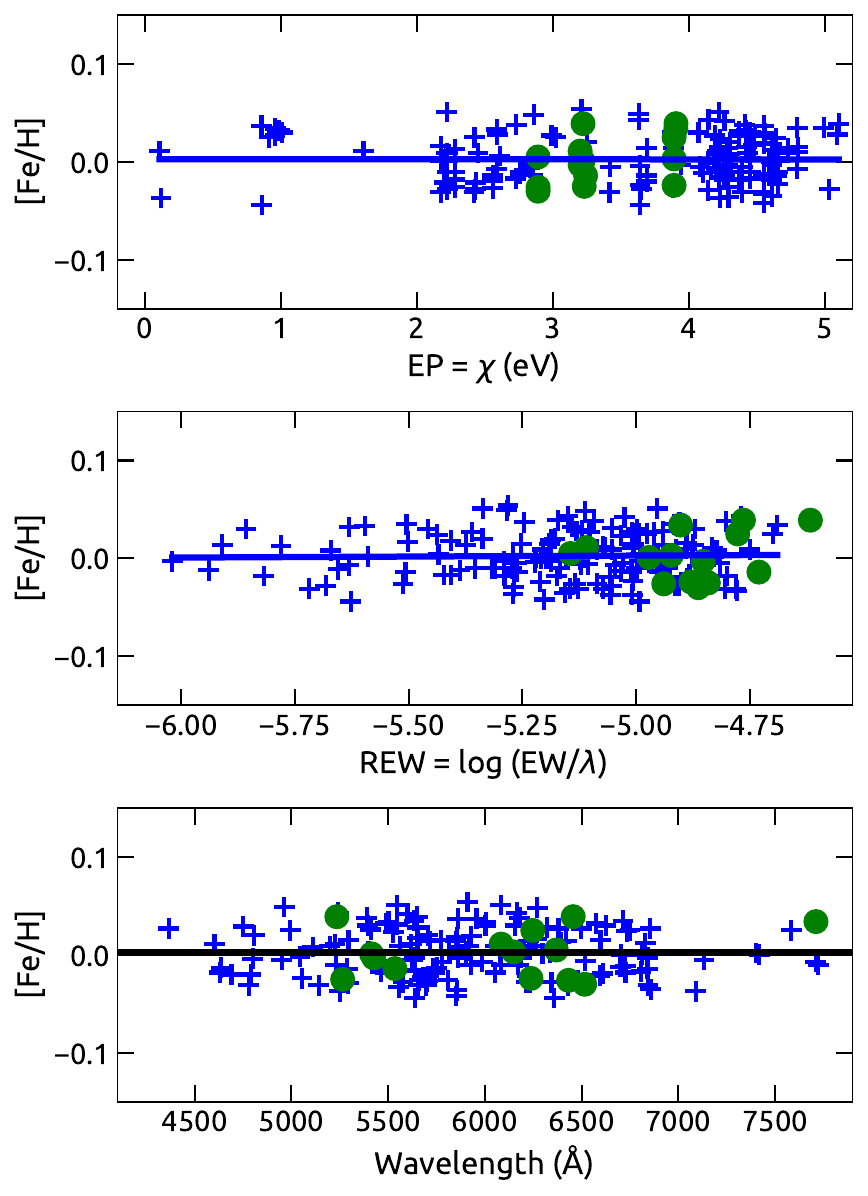}
\caption{Line-by-line iron abundance of HD 202772A, computed relative to HD 202772B, as a function of excitation potential (upper panel), reduced EW (middle panel), and wavelength (bottom panel). Blue crosses (green circles) correspond to Fe \textsc{i} (Fe \textsc{ii}) lines. In the top and middle panel, solid lines are linear fits to the Fe \textsc{i} data, while the solid black line in the bottom panel indicates the average iron abundance from all lines. \label{fig.equilibrium}}
\end{figure}

\begin{figure}
\centering
\includegraphics[width=0.40\textwidth]{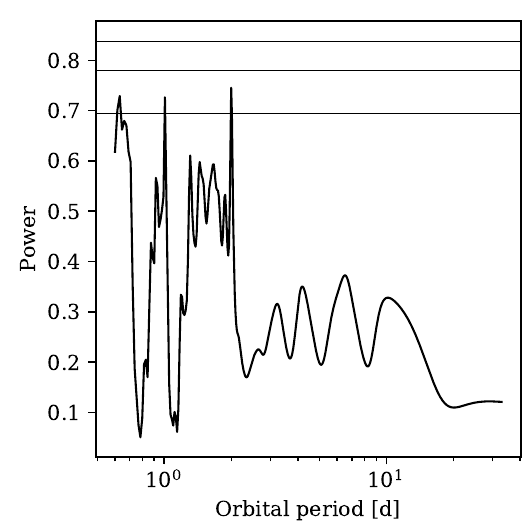}
\caption{Generalized Lomb Scargle periodogram of the model velocity residuals. The horizontal lines represent, from bottom to top, the estimate 10\%, 1\% and 0.1\% so-called False Alarm Probability, computed as the percentiles of the maximum power distribution of shuffled data (see text). None of the peaks appear significant. \label{fig.GLSomc}}
\end{figure}

\begin{table*}
\caption{Median values and 68\% confidence interval for the Physical and Orbital Parameters of HD~202772A~b}
\label{table.planetary.parameters}
\centering
\begin{tabular}{l c c}
\hline\hline
Parameter	&	Units	&	Best-fit solution	\\
\hline		
~~~~$P$\dotfill &Period (days)\dotfill &$3.30887750\pm0.00000051$\\ 
~~~~$T_{conj}$\dotfill &Time of conjunction  (BJD$_{\rm TDB}$)\dotfill 
&$2458328.68357\pm0.00016$\\
~~~~$T_S$\dotfill &Time of eclipse (BJD$_{\rm TDB}$)\dotfill &$2458330.332^{+0.034}_{-0.041}$\\
~~~~$a$\dotfill &Semi-major axis (au)\dotfill &$0.05208^{+0.00054}_{-0.00059}$\\
~~~~$R_P$\dotfill &Radius ($R_{\mathrm{J}}$)\dotfill &$1.627^{+0.070}_{-0.065}$\\
~~~~$M_P$\dotfill &Mass ($M_{\mathrm{J}}$)\dotfill &$1.018^{+0.066}_{-0.063}$\\
~~~~$e$\dotfill &Eccentricity \dotfill &$0.030^{+0.024}_{-0.019}$\\
~~~~$\omega_*$\dotfill &Argument of Periastron (Degrees)\dotfill &$99^{+46}_{-43}$\\
~~~~$e\cos{\omega_*}$\dotfill & \dotfill &$-0.003^{+0.016}_{-0.020}$\\
~~~~$e\sin{\omega_*}$\dotfill & \dotfill &$0.021^{+0.025}_{-0.018}$\\

~~~~$\rho_P$\dotfill &Density (cgs)\dotfill &$0.293^{+0.040}_{-0.035}$\\
~~~~$logg_P$\dotfill &Surface gravity (cgs) \dotfill &$2.978\pm0.041$\\
~~~~$T_{eq}$\dotfill &Equilibrium temperature (K)\dotfill &$2181^{+23}_{-21}$\\
~~~~$\Theta$\dotfill &Safronov Number \dotfill &$0.0378^{+0.0027}_{-0.0025}$\\
~~~~$\langle F \rangle$\dotfill &Incident Flux (10$^9$ erg s$^{-1}$ cm$^{-2}$)\dotfill &$5.13^{+0.21}_{-0.19}$ 
\smallskip\\\multicolumn{2}{l}{Primary transit parameters:}\smallskip\\					~~~~$R_P/R_*$\dotfill &Radius of planet in stellar radii \dotfill &$0.0655^{+0.0027}_{-0.0025}$\\
~~~~$a/R_*$\dotfill &Semi-major axis in stellar radii \dotfill &$4.384^{+0.084}_{-0.091}$\\
~~~~$i$\dotfill &Inclination (Degrees)\dotfill &$84.90^{+0.57}_{-0.55}$\\
~~~~$b$\dotfill &Transit Impact parameter \dotfill &$0.381^{+0.038}_{-0.042}$\\
~~~~$\delta_{\rm TESS}$\dotfill &Transit depth in TESS (fraction)\dotfill &$0.00467^{+0.00040}_{-0.00036}$\\
~~~~$T_{14}$\dotfill &Total transit duration (days)\dotfill &$0.23638^{+0.00063}_{-0.00062}$\\
~~~~$T_{FWHM}$\dotfill &FWHM transit duration (days)\dotfill &$0.21929\pm0.00035$\\
~~~~$\tau_S$\dotfill &Ingress/egress eclipse duration (days)\dotfill &$0.01802^{+0.0011}_{-0.00093}$\\
~~~~$P_T$\dotfill &A priori non-grazing transit prob \dotfill &$0.2179^{+0.0098}_{-0.0071}$\\
~~~~$P_{T,G}$\dotfill &A priori transit prob \dotfill &$0.2483^{+0.011}_{-0.0079}$\\
~~~~$u_{1}$\dotfill &linear limb-darkening coeff \dotfill &$0.195\pm0.017$\\
~~~~$u_{2}$\dotfill &quadratic limb-darkening coeff \dotfill &$0.306\pm0.025$\\
~~~~$A_D$\dotfill &Dilution from neighboring stars \dotfill &$0.167^{+0.066}_{-0.070}$

\smallskip\\\multicolumn{2}{l}{Radial velocity parameters:}\smallskip\\
~~~~$T_P$\dotfill &Time of Periastron (BJD$_{\rm TDB}$) \dotfill &$2458325.46^{+0.41}_{-0.37}$\\
~~~~$T_A$\dotfill &Time of Ascending Node (BJD$_{\rm TDB}$)\dotfill &$2458327.874^{+0.034}_{-0.022}$\\
~~~~$T_D$\dotfill &Time of Descending Node (BJD$_{\rm TDB}$)\dotfill &$2458326.176^{+0.023}_{-0.034}$\\
~~~~$K$\dotfill &RV semi-amplitude (m/s)\dotfill &$96.5^{+6.0}_{-5.7}$\\
~~~~$M_P\sin i$\dotfill &Minimum mass (M$_{\rm J}$)\dotfill &$1.014^{+0.066}_{-0.063}$\\
~~~~$M_P/M_*$\dotfill &Mass ratio \dotfill &$0.000566^{+0.000036}_{-0.000034}$\\
~~~~$d/R_*$\dotfill &Separation at mid transit \dotfill &$4.29^{+0.14}_{-0.19}$

\smallskip\\\multicolumn{2}{l}{Telescope Parameters:} \smallskip\\
~~~~$\gamma_{\rm CHIRON}$\dotfill &Relative RV Offset (m/s)\dotfill &$-2.7^{+4.6}_{-4.9}$\\ 
~~~~$\gamma_{\rm HARPS}$\dotfill &Relative RV Offset (m/s)\dotfill &$9\pm11$\\
~~~~$\gamma_{\rm TRES}$\dotfill &Relative RV Offset (m/s)\dotfill  &$98.6^{+8.5}_{-8.6}$\\
~~~~$\sigma_{J \rm CHIRON}$\dotfill &RV Jitter (m/s)\dotfill &$14.5^{+5.4}_{-4.0}$\\
~~~~$\sigma_{J \rm HARPS}$\dotfill &RV Jitter (m/s)\dotfill &$27.3^{+15}_{-8.3}$\\
~~~~$\sigma_{J \rm TRES}$\dotfill &RV Jitter (m/s)\dotfill &$23.7^{+10.}_{-8.2}$\\
~~~~$\sigma_{J^2 \rm CHIRON}$ \dotfill &RV Jitter Variance \dotfill &$210^{+190}_{-100}$ \\
~~~~$\sigma_{J^2 \rm HARPS}$ \dotfill &RV Jitter Variance \dotfill &$750^{+1100}_{-390}$\\
~~~~$\sigma_{J^2 \rm TRES}$ \dotfill &RV Jitter Variance \dotfill &$560^{+600}_{-320}$\\
\hline																									
\end{tabular}

\end{table*}


\bsp	
\label{lastpage}
\end{document}